\documentclass[11pt]{article}
\pdfoutput=1
\usepackage{jheppub}
\usepackage{amsmath,amssymb,amsfonts,graphicx,slashed,
color,amsthm,mathtools,upgreek, enumerate, 
tensor, subfig,hyperref,MnSymbol,tikz,
tkz-euclide,tikz-feynman,float,arydshln,bbm}
\hypersetup{colorlinks=true}

\usetikzlibrary{calc}

\allowdisplaybreaks

\newcommand{\beq}{\begin{equation}}
\newcommand{\eeq}{\end{equation}}
\newcommand{\beqn}{\begin{eqnarray}}
\newcommand{\eeqn}{\end{eqnarray}}

\newcommand{\be}{\begin{eqnarray}}
\newcommand{\ee}{\end{eqnarray}}
\newcommand{\bea}{\begin{align*}}
\newcommand{\eea}{\end{align*}}

\newcommand\numberthis{\addtocounter{equation}{1}\tag{\theequation}}
\numberwithin{equation}{section}

\newcommand{\CO}{{\cal O}}
\newcommand{\COt}{{\tilde{\cal O}}}
\def\<{\langle}
\renewcommand\>{\rangle}

\DeclareMathOperator{\csch}{csch}

\title{CFT$_2$ in the Bulk}
\author{Cameron V. Cogburn}

\affiliation{Department of Physics, Boston University \\ 590 Commonwealth Avenue, Boston, MA 02215, 
USA} 

\emailAdd{cogburn@bu.edu}

\abstract{We study non-Gaussian bulk 2d CFTs  in AdS$_2$ using boundary CFT techniques and recent 
results in JT/Schwarzian gravity. We highlight the constraints on the operator content of a theory imposed by 
the boundary conditions by examining the relation between correlator coefficients and OPE coefficients in 
the presence of a boundary. We then calculate bulk and boundary correlators for various boundary 
conditions. Schwarzian techniques are used to calculate gravitational correlators perturbatively in $1/c$.}

\begin{document}

\maketitle
\parskip=12pt

\section{Introduction}

Studying strongly coupled QFTs is difficult as perturbative calculational tools break down in the absence 
of a small expansion parameter. By ``strongly coupled" we mean the correlators of generic operators are  
non-Gaussian, i.e., they do not obey an expansion in a $``1/N"$ small parameter.  

AdS is a convenient background in which to study strongly coupled systems as it is a maximally 
symmetric space that inherently acts as a bulk IR regulator \cite{Callan:1989em}. One way to study 
strongly coupled QFTs in AdS non-perturbatively is to make use of a numerical scheme such as a lattice 
discretization \cite{Brower:2019kyh} or a Hamiltonian truncation \cite{Hogervorst:2021spa}. Another way is 
to use the property that a CFT in AdS$_d$ is related to its boundary CFT (BCFT) in the $d$-dimensional 
upper half-plane $\mathcal{H}^d \equiv \{\vec{x}+iy \, | \, y>0; \vec{x} \in \mathbbm{R}^{d-1} \, , y \in 
\mathbbm{R} \}$ via a Weyl transformation:
\be
ds^2_{\text{AdS}_d} = y^{-2}(dy^2 + d\vec{x}^2) = \Omega^2 ds^2_{\mathcal{H}^d} \; ,
\ee
with $\Omega = y^{-1}$. This has been utilized several times before \cite{Doyon:2004fv, Aharony:2010ay, 
Carmi:2018qzm, Paulos:2016fap, Herzog:2019bom, Aharony:2015hix, Giombi:2020rmc}; therefore we can 
use tools from the sizeable literature on BCFTs to study rational CFTs in AdS. 

Although the QFT is strongly coupled, gravity can be weakly coupled by inserting factors of the stress 
tensor into correlators, e.g., 
\be
\< \phi  \dots \phi T \dots T \> \; ,
\ee
where $\phi$ is a scalar and $T$ a stress tensor, the latter of which gravitons are coupled to. For a 
general non-Gaussian CFT we do not know these; similarly, in dimensions higher than two the $TT$ OPE 
structure is not unique \cite{Osborn:1993cr}. However, in CFT$_2$, adding factors of $T$ is controlled by 
conformal invariance, making such calculations tractable. Moreover, in 2d it is possible to use the dual 1d 
Schwarzian theory to calculate the gravitational correlators of a boundary quantity $\tilde{\CO}$,
\be
\< \tilde{\CO} \>_t = \int [Dt] \tilde{\CO} e^{-S[t]} \; , \quad S[t] \propto \int dt \, \{t,u\} \; ,
\ee
that would otherwise be calculated by inserting multiples of the stress tensor into a bulk correlator before 
taking it to the boundary. 

The purpose of this work is to show that in AdS$_2$ all the machinery exists to do perturbative 
gravitational calculations, and some types of non-perturbative ones, on strongly coupled conformal 
theories weakly coupled to gravity, using the unitary minimal models as an explicit example.  Section 
\ref{sec:CFT2AdS2} provides an overview of CFT$_2$ in AdS$_2$. We use the Ising model to explicitly 
show how boundary correlators can be obtained solely from the bulk-boundary OPE, and how boundary 
conditions constrain the bulk and boundary data. In Section \ref{sec:InfStrip}, by mapping to the infinite strip 
we can study the theory in the presence of two boundaries, such as in RS1 models. In Section 
\ref{sec:MMinGravity} we add gravity and calculate the gravitational correction for a generic minimal model $
\tilde{\phi}_{2,1}$ four-point function using Schwarzian techniques. Section \ref{sec:HigherDims} discusses 
the case of higher dimensions and how to calculate boundary gravitational corrections therein.
Finally, in Section \ref{sec:Conclusions} we conclude with a discussion.

\section{CFT$_2$ in AdS$_2$} \label{sec:CFT2AdS2}
We begin with a general discussion of putting a 2d CFT in AdS$_2$. Since AdS has a conformal boundary 
we need to understand how boundary conditions affect the bulk theory. There is an abundance of literature 
on CFTs in the presence of a boundary in flat space, from which we review the salient points and apply them 
to AdS. The main idea is that we can obtain boundary correlators solely from the bulk-boundary OPE, 
without needing to solve for the bulk correlators and then taking them to the boundary. We also review the 
constraints imposed from boundary conditions on the bulk correlators and the dual CFT$_1$ boundary 
operators. Boundary correlators are the main ingredient needed for the gravitational calculations of Section 
\ref{sec:MMinGravity}.

\subsection{Overview: boundary conditions} 
Many physical systems have some sort of boundary, so the study of conformal theories in the presence 
of a boundary has been well developed. The upper half-plane (UHP) is the prototypical space in which to 
study the consequences of demanding conformal invariance for a two-dimensional system in the presence 
of a boundary. A key point in the analysis is to use the ``doubling trick", in which a mirror image of the 
system is introduced in the lower half-plane and the now doubled system is analyzed in the full plane, 
$\mathbbm{R}^2$. Doing this in AdS$_2$ only introduces a Weyl factor. Upon taking the bulk correlators to 
the boundary we encounter a small puzzle, the clarification of which is a theme of Section 
\ref{sec:CFT2AdS2}.

We start by putting a CFT$_2$ in the UHP with a flat metric and a conformal boundary condition 
at the boundary $y=0$ by using the aforementioned doubling trick \cite{CARDY1984514}:
\be \label{eq:DoublingTrick}
\< \CO_1(z_1, \bar{z}_1) \dots \CO_n(z_n, \bar{z}_n) \>_{\rm UHP} = \< \CO_1(z_1) \CO_1(z^*_1) \dots 
\CO_n(z_n) 
\CO_n(z_n^*)\> \; .
\ee
Here $\CO_i(z)$ is the holomorphic half of the operator $\CO_i(z,\bar{z})$ and $z=x+i y$.  This construction 
explicitly preserves all of the symmetries (conformal transformations that leave the $x$-axis invariant as well 
as any holomorphic equations of motion for the $\CO$s). We will see later it also makes it convenient to set 
different boundary conditions by making different choices about what terms in the OPE of $\CO_i(z_i) 
\CO_i(z^*_i)$ are kept.   

To map this correlator to AdS$_2$, perform a Weyl transformation,
\be
g_{\mu\nu}(x,y) \rightarrow \Omega^2(x,y) g_{\mu\nu}(x,y) \; ,
\ee
with $\Omega(x,y) = y^{-1}$.   Under this transformation the CFT operators transform like 
$\Omega^{-\Delta}$, 
so we have
\begin{equation} \label{eq:CorrAdSFlat}
G^{(n)}_{\CO}(x_i, y_i) \equiv \< \CO_1(x_1, y_1) \dots \CO_n(x_n,y_n) \>_{{\rm AdS}_2} 
= y_1^{\Delta_1} \dots y_n^{\Delta_n} \< \CO_1(z_1) \CO_1(z^*_1) \dots \CO_n(z_n) \CO_n(z_n^*)\> \; .
\end{equation}
To obtain the boundary correlator, take the limit as the bulk operators approach the boundary:
\be
G^{(n)}_{\CO, \partial}(x_i) \equiv \< \tilde{\CO}_1(x_1) \dots \tilde{\CO}_n(x_n)\>_{\mathbb{R}}  
= \lim_{y_i \rightarrow 0} y_1^{-\tilde{\Delta}_1} \dots y_n^{-\tilde{\Delta}_n} G^{(n)}_{\CO}(x_i, y_i) \; .
\ee
Here, $\tilde{\Delta}_i$ are the dimensions of the dual CFT$_1$ operators.  By ``dual CFT$_1$ operator'' we 
mean the leading operator in the boundary limit of the bulk operator coming from the bulk-boundary operator 
expansion
\be \label{eq:BbOPE}
\CO_i(x,y) = \sum_{j} (2y)^{\Delta_{j} -\Delta_{i}} C^{a}_{ij} \tilde{\CO}_{j}(x) + \cdots
\ee
The $C^{a}_{ij}$ are the boundary OPE coefficients and the index $a$ signifies that the precise operators in 
the expansion are boundary condition-dependent. At the same time, because of the doubling trick, we can 
compare the bulk-boundary OPE (\ref{eq:BbOPE}) with the boundary limit of the $\CO_i \times \CO_i$ bulk 
OPE. Taking the bulk OPE
\be \label{eq:bulkOPEa}
\CO_i(z) \CO_i(w) &=& \sum_{\CO'} \frac{c_{ii \CO'}}{(z-w)^{2h_i - h_{\CO'}}} \CO'(w) 
\ee
in the boundary limit yields
\be \label{eq:bdryOPElimit}
\lim_{y\rightarrow 0} \CO_i(z) \CO_i(z^*) = \lim_{y\rightarrow 0} \CO_i(x+i y) \CO_i(x-i y) 
= \sum_{\CO'} c_{ii \CO'}(2y)^{- \Delta_i+  h_{\CO'}} \CO'(x)+ \dots
\ee
We see that $\tilde{\CO}$ is just the lowest-dimension operator $\CO'$ in the $\CO_i \times \CO_i$ OPE, 
and the dimension $\tilde{\Delta}$ of $\tilde{\CO}$ is $h_{\CO'}$.  

Usually the theory is defined so that the lowest-dimension operator in the $\CO_i \times \CO_i$ bulk OPE 
is the identity operator. However, we may have or want boundary conditions that necessitate having a 
vanishing identity contribution. By inspection of the above OPEs, we see that this means setting the OPE 
coefficients for all operators with $h_{\CO'} < 2h_i$ to zero, seemingly eliminating the identity as the dual 
CFT$_1$ operator. 

We can identify when this is the case by looking at the bulk-boundary OPE (\ref{eq:BbOPE}). The identity 
is not the dual CFT$_1$ operator when the bulk-boundary coupling coefficient to the identity $A^{a}_i \equiv 
C^{a}_{i\mathbbm{1}}$ vanishes. This can easily be read off from the modular matrix, as shown in the 
following subsection. The point is that we can obtain the boundary correlators solely from the bulk-boundary 
OPE, without needing to solve for bulk correlators and then taking them to the boundary. 

As an example, let us compute the boundary $2n$-point functions from bulk $\< \sigma \sigma \dots 
\sigma\>$ correlators of the Ising model. The $\sigma \times \sigma$ OPE is
\be
[\sigma] \times [\sigma] =  [1] + [\epsilon] \; .
\ee
where $[ \dots]$ indicates the full Virasoro irrep of operators. Pulling off the identity operator in order to 
impose Dirichlet boundary conditions, we find that the boundary correlators are
\be \label{eq:4ptepscorgen}
G^{(2n)}_{\sigma,\partial}(x_i) = \< \epsilon(x_1) \dots \epsilon(x_{2n})\> = \bigg| \text{Pf} \, \bigg[ 
\frac{1}{x_i 
- x_j} \bigg]_{1\leq i,j\leq 2n} \bigg|^2 \; ,
\ee
where $\text{Pf}$ is the Pfaffian. For example, with $n=2$ we have
\be \label{eq:4ptepscor}
G^{(4)}_{\sigma,\partial}(x_i) = \< \epsilon(x_1) \epsilon(x_2) \epsilon(x_3) \epsilon(x_4) \> = 
\frac{1}{x_{12} 
x_{34}} + \frac{1}{x_{13} x_{24}} + \frac{1}{x_{14} x_{23}} \; .
\ee
Although the Ising model is particularly simple, the above logic applies to any boundary correlator from a 
bulk 2d CFT.

Since the bulk-boundary identity coupling $A^{a}_i$ is dependent on the boundary condition $a$, so is 
the dual CFT$_1$ operator. Using Cardy boundary states we can understand precisely what boundary 
conditions correlate to what bulk-boundary coefficients, and hence to what dual CFT$_1$ operators.

\subsection{Boundary states}
As we saw above, the bulk-boundary OPE coefficients are dependent on the boundary condition, that is the 
physical boundary state. Since Cardy classified the allowed physical boundary states they are also known 
as Cardy boundary states. 

A Cardy boundary state $|\tilde{l}\>$ can be expanded in the basis of allowed states (Ishibashi states) 
$|j\rrangle$ corresponding to the bulk primary fields as \cite{Cardy:1989ir}
\be \label{eq:Sbasisexpansion1}
| \tilde{l} \> = \sum_{j} \frac{S_{ij}}{\sqrt{S_{0j}}} |j \rrangle \; ,
\ee
where $S$ is the modular matrix.\footnote{A more detailed discussion about the modular matrix, Cardy 
boundary states, and Ishibashi states can be found in Appendix \ref{app:ModMatrix}.} The matrix $S$ also 
encodes the coupling between bulk fields and the identity boundary operator through \cite{CARDY1991274}
\be \label{eq:onepointboundcoupling1}
A^{a}_j \equiv C^{a}_{j \mathbbm{1}} = \frac{ \llangle j | a \> }{ \llangle 0 | a \>} = \frac{S_{ja}}{S_{0a}} 
\left(\frac{S_{00}}{S_{j0}}\right)^{1/2} 
\; ,
\ee
where $j$ is an Ishibashi state and $a$ is a Cardy boundary state defined in (\ref{eq:Sbasisexpansion1}). 
This means that when $C^{a}_{j \mathbbm{1}} = 0$ the identity is \textit{not} the leading operator in the 
expansion of the associated bulk operator $\phi_j$ in terms of its dual boundary operators in the 
short-distance expansion (\ref{eq:BbOPE}). For $C^{a}_{j \mathbbm{1}} = 0$ we need the matrix element 
$S_{ja} = 0$, which can be calculated from (\ref{eq:Selements}). Since Dirichlet (free) boundary conditions 
correspond to requiring there being no identity on the boundary, we can identify the boundary conditions 
corresponding to the free states by identifying the vanishing $S$-matrix elements.

\subsection{Boundary data (dual CFT$_1$ operators)} \label{subsec:dualops}

Because the boundary conditions determine the bulk-boundary coefficients $C^{a}_{ij}$ they also specify the 
dual CFT$_1$ operators. For the Ising case, expanding $\sigma$ gives
\begin{align*}
\sigma(z) &= (2y)^{\tilde{\Delta}_1 - \Delta_{\sigma}} C^{a}_{\sigma \mathbbm{1}} \tilde{\mathbbm{1}} + 
(2y)^{\tilde{\Delta}_{\epsilon} - 
\Delta_{\sigma}} C^{a}_{\sigma \epsilon} \tilde{\epsilon}(x) + \cdots \\
&= \frac{1}{(2y)^{1/8}} C^{a}_{\sigma \mathbbm{1}} \tilde{\mathbbm{1}} +  (2y)^{3/8} C^{a}_{\sigma 
\epsilon} \tilde{\epsilon}(x) + \cdots
\numberthis 
\end{align*}
Doing the same for $\epsilon$ gives
\begin{align*}
\epsilon(z) &= (2y)^{\tilde{\Delta}_1 - \Delta_{\epsilon}} C^{a}_{\epsilon \mathbbm{1}} \tilde{\mathbbm{1}} 
+ (2y)^{\tilde{\Delta}_{\sigma} - \Delta_{\epsilon}} C^{a}_{\epsilon \sigma} \tilde{\sigma}(x)
+ \cdots \\
&= \frac{1}{(2y)^{1/2}} C^{a}_{\epsilon \mathbbm{1}} \tilde{\mathbbm{1}} 
+ \frac{1}{(2y)^{3/8}} C^{a}_{\epsilon \sigma} \tilde{\sigma}(x)
+ \cdots
\numberthis 
\end{align*}
The bulk-boundary structure constants for the different boundary conditions in the Ising model are calculated 
in (\ref{eq:bulkboundconst1}) - (\ref{eq:bulkboundconst4}). Applying the results here, for the fixed $(\pm)$ 
case 
we get:
\begin{align*} \label{eq:IsingCFT1OpsFixed}
\sigma(z) = \frac{\pm 2^{1/4}}{(2y)^{1/8}} \tilde{\mathbbm{1}} + \cdots \; , \quad \epsilon(z) = 
\frac{1}{(2y)^{1/2}} 
\tilde{\mathbbm{1}} + \cdots \; . 
\numberthis
\end{align*}
Whereas for the free $(f)$ case we get:
\begin{align*} \label{eq:IsingCFT1OpsFree}
\sigma(z) = \frac{1}{(2y)^{3/8}} \left( \sqrt{2} \alpha^{ff}_{\tilde{\epsilon}}   \right)^{-1/2}  \tilde{\epsilon}(x) + 
\cdots \; , \quad \epsilon(z) = 
-\frac{1}{(2y)^{1/2}} 
\tilde{\mathbbm{1}} + \cdots \; . \numberthis
\end{align*}
This shows that for $(\pm)$ boundary conditions, the dual CFT$_1$ operator for both $\sigma$ and 
$\epsilon$ is simply the boundary identity $\tilde{\mathbbm{1}}$. For $(f)$ boundary conditions, the dual 
CFT$_1$ operator for $\epsilon$ is again $\tilde{\mathbbm{1}}$ but for $\sigma$ it is the dual energy density 
operator $\tilde{\epsilon}$. 

We saw that for the Ising model the form of the dual CFT$_1$ operators were determined by the 
bulk-boundary structure constants $C^{a}_{j i}$, which in turn were dependent on the modular matrix through 
(\ref{eq:onepointboundcoupling1}). The generalization to other minimal models is straightforward: given a 
bulk minimal model and boundary conditions we can apply 
(\ref{eq:onepointboundcoupling1}) to determine the bulk-boundary structure constants, and hence the 
allowed dual operator spectrum.

\subsection{Bulk data from boundary conditions}

We can see how the boundary conditions also set the form of the bulk correlators. Let us restrict the bulk 
CFT$_2$ theory in AdS$_2$ to the unitary minimal models. Our first goal is to calculate a bulk two-point 
function $G^{(2)}_{\phi_1 \phi_2}(z_1,\bar{z}_1,z_2,\bar{z}_2)=\< \phi_{h_1}(z_1, \bar{z}_1) \phi_{h_2}(z_2, 
\bar{z}_2) \>$ in this geometry. Here, $\phi_{h_i}$ is a primary scalar field with dimension $h_i$. Because of 
the doubling trick (\ref{eq:DoublingTrick}), the two-point function will be a linear combination of the same 
conformal blocks as the associated four-point function in the full space produced by the doubling:
\be \label{eq:4scalar}
G^{(4)}_{\phi_1 \phi_2 \phi_1 \phi_2}(z_1, z_2, z_3, z_4) = \< \phi_{h_1}(z_1) \phi_{h_2}(z_2) 
\phi_{h_1}(z_3) 
\phi_{h_2}(z_4) \> \; ,
\ee
with $z_3 = z^{*}_1$ and $z_4 = z^{*}_2$. The process of projecting out null states to obtain correlator 
differential equations is famously well known, as solving the appropriate differential equation amounts to a 
complete determination of the correlators in a minimal theory \cite{DiFrancesco:639405}. 

As a specific example, let us study the Ising spin two-point function $\< \sigma \sigma \>$ in AdS$_2$ 
by putting it in the Poincar\'{e} UHP. From (\ref{eq:CorrAdSFlat}) this is equal to a Weyl transformed full-plane four-point function\footnote{We have intentionally absorbed a factor of $\sqrt{-1}$ into the second term so that $c_2^{\sigma \sigma}$ is real. This is because the cross-ratio $x$ can be written as
\be
x = \frac{z_{12} z_{34}}{z_{13}z_{24}} = - \frac{\rho^2 + (y_1 - y_2)^2}{4 y_1 y_2}  \; ,
\ee
where $\rho \equiv x_2 - x_1$, meaning the physical region is given by $x < 0$. We have also picked a 
convenient overall constant $2^{-1/4}$ to keep the relation between correlator and OPE coefficients simple.}
\be \label{eq:G2ptcoeffs1}
G^{(2)}_{^{\sigma}} =  \frac{1}{2^{1/4}(y_1 y_2)^{1/8}} \frac{1}{[x(x-1)]^{1/8}} \left( c^{\sigma \sigma}_1 
\frac{\sqrt{\sqrt{1-x}+1}}{\sqrt{2}} 
+ c^{\sigma \sigma}_2 \sqrt{2} \sqrt{\sqrt{1-x}-1} \right) \; .
\ee
The coefficients $c^{\sigma \sigma}_1$, $c^{\sigma \sigma}_2$ are determined from the boundary 
conditions, and could be found at this point from the asymptotic behavior of the correlator near the real-axis 
boundary.

At the same time, the bulk OPE gives the form
\be \label{eq:sigmabulktwo}
G^{(2)}_{^{\sigma}} = C_{\sigma \sigma \mathbbm{1}} |z_{12}|^{-1/4} + C_{\sigma \sigma \epsilon} 
A_{\epsilon} |z_{12}|^{1/4} \; ,
\ee
with the bulk coefficients given by $C_{\sigma \sigma \mathbbm{1}} = 1$, $A_{\epsilon} = 1$, and 
$C_{\sigma \sigma \epsilon} = \pm \frac{1}{2}$, the sign being determined by the boundary condition.\footnote{ The ``extraordinary transition" gives the plus sign whereas the ``ordinary transition" gives the minus sign.}
Taking the limits $x \rightarrow 0$ and $x \rightarrow -\infty$ in (\ref{eq:G2ptcoeffs1}) 
and equating with (\ref{eq:sigmabulktwo}) sets 
\be
c_{1}^{\sigma \sigma} = C_{\sigma \sigma \mathbbm{1}} = 1 \; ,
\quad \quad 
c_{2}^{\sigma \sigma} = C_{\sigma \sigma \epsilon} A_{\epsilon}  = \pm\frac{1}{2} \;.
\ee
What we have reviewed is that the physical boundary conditions change the OPE coefficients and the 
conformal block coefficients; or equivalently, an appropriate change in the OPE coefficients induces a 
change in the boundary conditions.

\section{Two Boundaries: The RS1 Setup} \label{sec:InfStrip}
Suppose we are interested in analyzing the case of a CFT$_2$ in the presence of two boundaries in 
AdS$_2$. Such a situation arises, for example, if we have a Randall-Sundrum 1 (RS1) model with two 
branes \cite{Randall:1999ee}. In this model there is a Planck brane at the conformal boundary $y=0$ in the 
UHP and a TeV brane at a distance $y=L$. Multiple boundaries are also useful, for example, in describing 
the RG flow between theories using a RG brane \cite{Gaiotto:2012np}.

A convenient geometry to analyze AdS$_2$ with two boundaries is the conformally equivalent infinite 
strip with finite width $L$. The strip geometry is obtained from the UHP through the conformal 
map
\be
w = \frac{L}{\pi} \ln z \; ,
\ee
where $w = u + iv$ and $z = x + iy$ are the respective holomorphic coordinates for the strip and for the 
UHP. Note how the UHP positive real axis $(x>0)$ is mapped to the lower edge of the strip $(v=0)$ and the 
negative real axis $(x<0)$ to the upper edge $(v=L)$. Therefore, we can obtain a strip with different 
boundary conditions on its two edges by looking at the UHP with differing boundary conditions along its positive and negative real axis, mediated by a boundary condition-changing operator at the origin. We again using the Ising model to illustrate calculations.

\subsection{Boundary correlators}
The analysis for the bulk spin two-point function in the flat half-plane metric is done in 
\cite{DiFrancesco:639405}. For correlators in AdS there is a Weyl factor to go from the flat metric to the 
Poincar\'{e} one. Boundary correlators on the strip are obtained by taking $v \rightarrow 0$. 

First consider the case where the two boundaries have the same boundary condition. A two-point function 
on the strip is given by
\begin{align*}
\< \CO_1(w_1, \bar{w}_1) \CO_2(w_2, \bar{w}_2) \>_{\text{strip}} =  
\prod_{i=1}^{2} \left(\frac{dz_i}{dw_i}\right)^{h_i} \left(\frac{d\bar{z}_i}{d\bar{w}_i}\right)^{\bar{h}_i}
\Omega^{-\Delta_{i}}_i  
\< \CO_1(z_1) \CO_2(z_2) \CO_1(z^{*}_1) \CO_2(z^{*}_2) \> \; , \numberthis
\end{align*}
where the doubling trick has been applied and the bulk four-point function pre-factor
\be
\prod_{i=1}^{2} \left(\frac{dz_i}{dw_i}\right)^{h_i} \left(\frac{d\bar{z}_i}{d\bar{w}_i}\right)^{\bar{h}_i}
\Omega^{-\Delta_{i}}_i = 
\bigg[ \Big(\frac{\pi}{L} \Big)^2 e^{\frac{\pi u_1}{L}}  v_1^2  \bigg]^{h_1} 
\bigg[ \Big(\frac{\pi}{L} \Big)^2 e^{\frac{\pi u_2}{L}}  v_2^2 \bigg]^{h_2} \; ,
\ee
comes from the Weyl transformation from the Poincar\'{e} UHP to the flat UHP and then to the 
strip. The boundary correlator can be obtained by going to the boundary via 
\be
\< \CO_1(w_1, \bar{w}_1) \CO_2(w_2, \bar{w}_2) \>_{\text{strip},\,\partial}^{(a)} = \lim_{v_i \rightarrow 0} 
v^{-\Delta_{\tilde{\CO}_1^{(a)}}}_1 
v^{-\Delta_{\tilde{\CO}_2^{(a)}}}_2 \< \CO_1(w_1, \bar{w}_1) \CO_2(w_2, \bar{w}_2) \>_{\text{strip}} \; ,
\ee
where the superscript notation $(a)$ indicates that the dual CFT$_1$ operator is boundary 
condition-dependent. 

Let us once again use the Ising spin correlator as an example. For free $(f)$ boundary conditions 
(ordinary case), the dual boundary operator from (\ref{eq:IsingCFT1OpsFree}) is $\tilde{\epsilon}$ with 
scaling dimension $\Delta_{\tilde{\epsilon}} = 1/2$. This gives for the strip boundary spin-spin correlator
\be
\< \sigma(w_1, \bar{w}_1) \sigma(w_2, \bar{w}_2) \>_{\text{strip},\,\partial}^{(ff)} = \frac{\pi}{2^{1/4} L} 
\csch \frac{\pi u}{2L} \; ,
\ee
where $u \equiv u_1 - u_2$. For fixed $(\pm)$ boundary conditions (extraordinary case) the dual CFT$_1$ 
operator from (\ref{eq:IsingCFT1OpsFixed}) is the identity, $\Delta_{\tilde{\mathbbm{1}}} = 0$. Taking the 
correlator to the edge yields
\begin{align*}
&\lim_{v_i \rightarrow 0} v_1^{0} v_2^{0} \< \sigma(w_1, \bar{w}_1) \sigma(w_2, \bar{w}_2) 
\>_{\text{strip}}^{(\pm \pm)} = \\
&2^{3/4} + \frac{\pi^2}{24 \, 2^{1/4} L^2}(v_1^2 + v_2^2) + \frac{7\pi^4}{3840 \, 2^{1/4} L^4}(v_1^4 + 
v_2^4) + v_1^2 v_2^2 
\left( \frac{\pi^2(1+36 \csch^4 \frac{\pi u}{2L})}{1152 \times 2^{1/4} L^4}  \right) \\
&= C_{0} +  v_1^2 v_2^2 C_{2} + \cdots  \numberthis
\end{align*}
As expected, the first term is the boundary operator $\tilde{\mathbbm{1}}$ and has a constant coefficient 
$C_0 = 2^{3/4}$. The next lowest dimension operator is the dimension-2 ``stress tensor" with coefficient $C_2$.

We can repeat the preceding steps to calculate the other Ising two-point correlators on the strip, giving the 
results
\begin{align*}
\< \epsilon(w_1, \bar{w}_1) \epsilon(w_2, \bar{w}_2) \>_{\mathrm{strip}, \, \partial}^{(ff) \,  \mathrm{or} \, 
(\pm \pm)} &= 
\frac{1}{4}  + \frac{\pi^2}{24 L^2}(v_1^2 + v_2^2) + \frac{7\pi^4}{1440 L^4}(v_1^4 + v_2^4) + v_1^2 v_2^2 
\left( \frac{\pi^4( 1+36 
\csch^4 \frac{\pi u}{2L})}{144 \times L^4}  \right) \\
&= D_{0} +  v_1^2 v_2^2 D_{2} + \cdots \numberthis
\end{align*}
and
\begin{align*}
\< \epsilon(w_1, \bar{w}_1) \sigma(w_2, \bar{w}_2) \>_{\text{strip}}^{(\pm \pm)} &= 
\left(\frac{\pi}{128 \, L} \right)^{1/8} + \cdots + v_1^2 v_2^2 \left( \frac{\pi^{33/8}( 1+36 \csch^4 \frac{\pi 
u}{2L})}{288 
\times 2^{7/8} 
L^{33/8}}  \right) \\
&= E_{0} +  v_1^2 v_2^2 E_{2} + \cdots \numberthis 
\end{align*}
Similar to before, the first term is the boundary identity operator with constant coefficient and the next 
term a dimension-2 operator.

\subsection{Energy spectrum}
Through the boundary correlators we can write down the spectrum of the boundary theory. This is 
particularly useful if our setup is a RS1 model where we would like to know the spectrum of particles on the 
boundary. The boundary spectrum can easily be read off by expanding the boundary correlator
\be
\< \tilde{\CO}(u_1) \tilde{\CO}(u_2) \> = \sum_{n} | \< 0 | \CO | n\>|^2 e^{-\Delta E_nu} \; ,
\ee 
where $\Delta E_n = E_n - E_0$ gives the energy of the particle above the ground state and $|c_n|^2 = | 
\< 0 | \CO | n\>|^2$ is the expansion coefficient. Doing this for the Ising spin-spin correlator in the ordinary 
transition gives
\begin{align*}
\< \sigma \sigma \>_{\text{strip}}^{\mathrm{(ff)}} = \frac{\pi}{2^{1/4} L} \csch \frac{\pi u}{2L} &= 
\frac{\pi}{2^{1/4} 
L} \frac{2}{e^{\frac{\pi 
u}{2L} }(1+e^{-\frac{\pi u}{L} })}\\
&= \frac{2^{3/4}\pi}{L} \left( e^{-\frac{\pi u}{2L} } + e^{-\frac{3\pi u}{2L} } + e^{-\frac{5\pi u}{2L} } + \cdots 
\right) \; . \numberthis
\end{align*}
For fixed $(\pm)$ boundary conditions we have to look at the term $C_2$ to expand. After normalizing by 
the short-distance two-point function we get
\be 
2\times 2^{1/4} \times C_2  =
\frac{\pi^4}{L^4} \left( \frac{1}{576}  + e^{-\frac{2\pi u}{L}} + 4e^{-\frac{3\pi u}{L}} + 10e^{-\frac{4\pi u}{L}} + 
\cdots   \right) \; .
\ee
For the energy-energy correlator, we do the same for the term $D_2$ to get
\be
D_2  =
\frac{\pi^4}{L^4} \left( \frac{1}{576}  + e^{-\frac{2\pi u}{L}} + 4e^{-\frac{3\pi u}{L}} + 10e^{-\frac{4\pi u}{L}} + 
\cdots   \right) \; ,
\ee
the same as the extraordinary spin-spin case. 
Finally, for the energy-spin correlator, looking at the term $E_2$ gives
\be
\left( \frac{L}{2\pi} \right)^{1/8} E_2  =
\frac{\pi^4}{L^4} \left( \frac{1}{576}  + e^{-\frac{2\pi u}{L}} + 4e^{-\frac{3\pi u}{L}} + 10e^{-\frac{4\pi u}{L}} + 
\cdots   \right) \; .
\ee
It is interesting to note that $\<\sigma \sigma\>^{(ff)}_{\mathrm{strip}}$ is the least suppressed by the 
system size $L$, 
compared to $\<\sigma \sigma\>^{(\pm)}_{\mathrm{strip}}$ or the $\< \epsilon \epsilon 
\>_{\mathrm{strip}}$ 
and $\< \sigma \epsilon 
\>_{\mathrm{strip}}$.
Also note that the amplitude for the mixed correlator $\< \sigma \epsilon \>$ is suppressed by an extra 
factor of $L^{-1/8}$ compared to the fixed $\< \sigma \sigma \>$ and $\<\epsilon \epsilon \>$ ones.

\subsubsection{Two-point functions with differing boundary conditions}
Generically, the two boundary conditions can be different. To induce a change of boundary conditions, a 
boundary condition-changing operator can be inserted at the origin on the UHP \cite{Cardy:1989ir}. 

How to calculate Ising correlators with differing boundary conditions on the infinite strip in flat space can be 
found in \cite{BURKHARDT1991653}. 
A key relation is that the conformal Ward identity in the 
presence of mixed boundary conditions $ab$ on the UHP links the $n$-point correlator 
in this mixed boundary, half-plane geometry with the bulk $(2n+2)$-point function:
\begin{align*}
\< \CO_1(z_1, \bar{z}_1) \dots \CO_n(z_n, \bar{z}_n) \>_{a b} = \lim_{x_1 \rightarrow 0 \atop {x_2 
\rightarrow -\infty}} (x_1 - x_2)^{2 
h_{a b}} \< \CO_1(z_1) \CO_1(z^{*}_1) \dots \CO_n(z_n)  \CO_n(z^{*}_n)\, \CO_{a b}(x_1) \CO_{a b}(x_2) \> \; , \numberthis
\end{align*}
where the $x_i$ are real coordinates on the boundary, and the $z_i$ are complex coordinates in the bulk.
Therefore, the same sequence of manipulations are done as in the case where the boundary 
conditions are the same, just now we need to calculate six-point functions instead of four-point functions. 
Again we use the Ising model as an explicit example.

The spin-spin boundary correlator between fixed boundary conditions $(+-)$, where the boundary 
condition-changing operator is given by the boundary operator $\tilde{\epsilon}$, is found to be
\begin{align*}
\< \sigma(1) \sigma(2)  \>_{\mathrm{strip}, \, \partial}^{(+-)} &= 
1 - \frac{23 \pi^2}{48 \, L^2}(v_1^2 + v_2^2) + \frac{247 \pi^4}{7680 \, L^4}(v_1^4 + v_2^4) \\
&+ v_1^2 v_2^2  \frac{\pi^4}{18432 \, L^4} \left( (-429 + 188 \cosh \frac{\pi u}{L} + 529 \cosh \frac{2 \pi 
u}{L}) 
\csch^4 \frac{\pi u}{2L}  
\right) \\
&= F_{0} +  v_1^2 v_2^2 F_{2} + \cdots  \numberthis
\end{align*}
The first term is the boundary identity operator and has a constant coefficient $F_0 = 1$ indicating the 
normalization. The next lowest dimension operator is again the dimension-2 stress tensor term. Continuing 
with the other two-point functions give
\begin{align*}
\< \sigma(1) \sigma(2) \>_{\mathrm{strip}, \, \partial}^{(+f)} = 
&1 - \frac{\pi^2}{24 \, L^2}(v_1^2 + v_2^2) - \frac{\pi^4}{960 \, L^4}(v_1^4 + v_2^4) \\&+ v_1^2 v_2^2 
\frac{\pi^4}{4608 \, L^4} \left( (3 + 
68 \cosh \frac{\pi u}{L} + \cosh \frac{2 \pi u}{L}) \csch^4 \frac{\pi u}{2L}  \right) \\
&= G_{0} +  v_1^2 v_2^2 G_{2} + \cdots  \numberthis
\end{align*}
\begin{align*}
\< \epsilon(1) \epsilon(2) \>_{\mathrm{strip}, \, \partial}^{(+-)} = 
&\frac{1}{4} - \frac{23\pi^2}{24 \, L^2}(v_1^2 + v_2^2) + \frac{247 \pi^4}{1440 \, L^4}(v_1^4 + v_2^4)  \\ 
&+v_1^2 v_2^2 
\frac{\pi^4}{1152 
\, L^4} \left( (-429 + 188 \cosh \frac{\pi u}{L} + 529 \cosh \frac{2 \pi u}{L}) \csch^4 \frac{\pi u}{2L}  \right) \\
&= H_{0} +  v_1^2 v_2^2 H_{2} + \cdots  \numberthis
\end{align*}
\begin{align*}
\< \epsilon(1) \epsilon(2) \>_{\mathrm{strip}, \, \partial}^{(+f)} = 
&\frac{1}{4} - \frac{\pi^2}{12 \, L^2}(v_1^2 + v_2^2) + \frac{ \pi^4}{180 \, L^4}(v_1^4 + v_2^4)  \\ &+v_1^2 
v_2^2 \frac{\pi^4}{288 \, L^4} 
\left( (3 + 68 \cosh \frac{\pi u}{L} +  \cosh \frac{2 \pi u}{L}) \csch^4 \frac{\pi u}{2L}  \right) \\
&= I_{0} +  v_1^2 v_2^2 I_{2} + \cdots  \numberthis
\end{align*}
\begin{align*}
\< \sigma(1) \epsilon(2) \>_{\mathrm{strip}, \, \partial}^{(+-)} &= 
\frac{\pi^{1/8}}{2 \, L^{1/8}} \bigg [ 1 - \frac{23 \pi^2}{48 \, L^2}(v_1^2 + 184v_2^2) + \frac{ 247 \pi^4}{7680 
\, L^4}(v_1^4 + \frac{64}{3} 
v_2^4) \\ &+v_1^2 v_2^2 \frac{\pi^4}{2304 \, L^4} \left( (-429 + 188 \cosh \frac{\pi u}{L} +  529 \cosh \frac{2 
\pi u}{L}) \csch^4 \frac{\pi 
u}{2L}  \right) \bigg ] \\
&= J_{0} +  v_1^2 v_2^2 J_{2} + \cdots  \numberthis
\end{align*}
\begin{align*} 
\< \sigma(1) \epsilon(2) \>_{\mathrm{strip}, \, \partial}^{(+f)} &= 
\frac{\pi^{1/8}}{ L^{1/8}} \bigg [ 1 - \frac{ \pi^2}{24 \, L^2}(v_1^2 + 8v_2^2) + \frac{  \pi^4}{960 \, L^4}(v_1^4 
+ \frac{64}{3} v_2^4)  \\ 
&+ v_1^2 v_2^2 \frac{\pi^4}{576 \, L^4} \left( (3 + 68 \cosh \frac{\pi u}{L} +   \cosh \frac{2 \pi u}{L}) \csch^4 
\frac{\pi u}{2L}  \right) \bigg ] 
\\
&= K_{0} +  v_1^2 v_2^2 K_{2} + \cdots  \numberthis
\end{align*}
By expanding the stress tensor term, we can find the energy spectrums as well:
\be
F_2  =
\frac{\pi^4}{L^4} \left( \frac{529}{2304}  + e^{-\frac{\pi u}{L}} + \frac{9}{4}  e^{-\frac{2\pi u}{L}} + 4 
e^{-\frac{3\pi 
u}{L}} + \cdots   \right)
\ee
\be
G_2  =
\frac{\pi^4}{L^4} \left( \frac{1}{576}  + \frac{1}{8} e^{-\frac{\pi u}{L}} + \frac{1}{2} e^{-\frac{2\pi u}{L}} + 
\frac{11}{8} e^{-\frac{3\pi u}{L}} + 
\cdots   \right)
\ee
\be
H_2  =
\frac{\pi^4}{L^4} \left( \frac{529}{144}  + 16 e^{-\frac{\pi u}{L}} + 36 e^{-\frac{2\pi u}{L}} + 64 e^{-\frac{3\pi 
u}{L}} + \cdots   \right)
\ee
\be
I_2  =
\frac{\pi^4}{L^4} \left( \frac{1}{36}  + 2 e^{-\frac{\pi u}{L}} + 8 e^{-\frac{2\pi u}{L}} + 22 e^{-\frac{3\pi u}{L}} + 
\cdots   \right)
\ee
\be
J_2  =
\frac{\pi^4}{L^4} \left( \frac{529}{288}  + 8 e^{-\frac{\pi u}{L}} + 18 e^{-\frac{2\pi u}{L}} + 32 e^{-\frac{3\pi 
u}{L}} 
+ \cdots   \right)
\ee
\be
K_2  =
\frac{\pi^4}{L^4} \left( \frac{1}{72}  +  e^{-\frac{\pi u}{L}} + 4 e^{-\frac{2\pi u}{L}} + 11 e^{-\frac{3\pi u}{L}} + 
\cdots   \right)
\ee

Although the above calculations have been done explicitly with the Ising model, generalization to other 
minimal models is straightforward. Given any minimal mode we can deduce the dual CFT$_1$ operators 
using the logic of Sec.~\ref{subsec:dualops}. Knowing these, we can calculate the boundary correlators 
on the infinite strip and read off the spectrum.

\section{Gravitational Corrections} \label{sec:MMinGravity}

\subsection{Gravitational coupling to a non-Gaussian theory} \label{sec:GravCoupling}

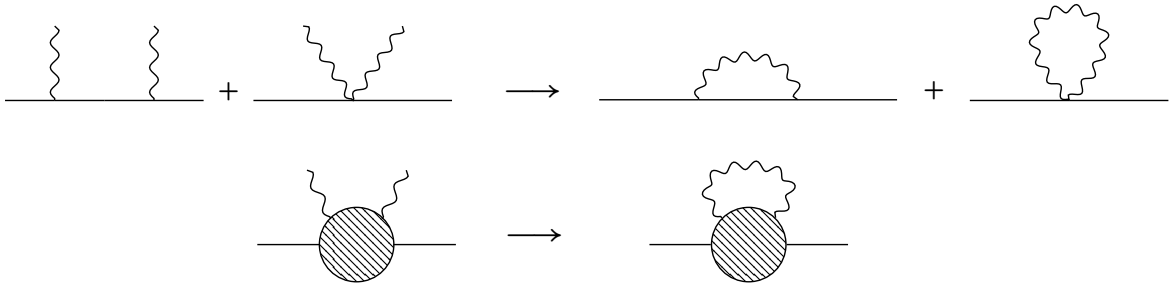
\begin{figure}[t] 
  \centering
  \hspace{-2.5cm}
    \scalebox{1.32}{
    \begin{tikzpicture}[scale=1.0]
  \def\leglength{1}
  \begin{feynman}
    \vertex (o) at (0,0);
    \vertex (a) at (-1.0,0);
    \vertex (b) at ( -0.5,0);
    \vertex (c) at (-0.5, 0.75);
    \vertex (d) at ( 0.5, 0);
    \vertex (e) at ( 0.5, 0.75);
    \vertex (f) at ( 1.0, 0);
    \diagram* {
      (a) -- (b) -- (o) -- (d) -- (f),
      (b) -- [boson] (c),
      (d) -- [boson] (e),
    };
  \end{feynman}
\end{tikzpicture}
\( + \)
\begin{tikzpicture}[scale=1.0]
  \def\leglength{1}
  \begin{feynman}
    \vertex (o) at (0,0);
    \vertex (a) at (-1.0,0);
    \vertex (b) at ( 0,0);
    \vertex (c) at (-0.5, 0.75);
    \vertex (d) at ( 0, 0);
    \vertex (e) at ( 0.5, 0.75);
    \vertex (f) at ( 1.0, 0);
    \diagram* {
      (a) -- (b) -- (o) -- (d) -- (f),
      (b) -- [boson] (c),
      (d) -- [boson] (e),
    };
  \end{feynman}
\end{tikzpicture}
\(  \quad \longrightarrow \)
\begin{tikzpicture}[scale=1.0]
\tikz{\begin{feynman}
\diagram*{
a -- b -- c -- d,
b -- [boson, half left] c;
};\end{feynman}}
\hspace{0.25cm}
\( + \) 
\feynmandiagram [horizontal=a to c, node distance=5.0cm, baseline=0] {
  a -- [dot] o -- [boson, out=135, in=45, loop, min distance=1.75cm] o -- c,
};
\end{tikzpicture}}
  \newline \newline
  \scalebox{1.32}{
\begin{tikzpicture}[scale=1.0]
\centering
\hspace{-1.27cm}
  \def\leglength{1}
  \begin{feynman}
    \vertex[blob] (m) at (0, 0) {};
    \vertex (o1) at (-0.37,0);
    \vertex (o2) at (0.37,0);
    \vertex (a) at (-1.0,0);
    \vertex (b) at ( -0.26,0.26);
    \vertex (c) at (-0.5, 0.75);
    \vertex (d) at ( 0.26, 0.26);
    \vertex (e) at ( 0.5, 0.75);
    \vertex (f) at ( 1.0, 0);
    \diagram* {
      (a) -- (o1),
      (o2) -- (f),
      (b) -- [boson] (c),
      (d) -- [boson] (e),
    };
  \end{feynman}
\hspace{1.5cm}
\( \quad \quad \longrightarrow \)
\hspace{0.86cm}
\feynmandiagram [horizontal=a to c, node distance=5.0cm,  baseline=0] {
  a -- [dot] o [blob] -- [boson, out=135, in=45, loop, min distance=1.0cm] o -- c
};
\end{tikzpicture}  }
  \caption{\textit{Top:} In a weakly coupled theory, gravitational soft modes become loop corrections that 
can be calculated using perturbative techniques such as propagators and vertices.
  \textit{Bottom:} In a strongly coupled theory perturbative techniques break down. To calculate 
gravitational loop corrections we have to integrate matter correlators with insertions of the stress tensor. 
This can equivalently be done using Schwarzian techniques.} \label{fig:WeakVsStrong}
\end{figure}

Now we weakly couple gravity to the strongly coupled CFT$_2$ and calculate gravitational 
corrections. For a Gaussian bulk theory, perturbation theory allows gravitational contributions to 
be calculated from the Feynman rules (cf.~Fig.~\ref{fig:WeakVsStrong}). However, in a strongly coupled 
theory, normal perturbative techniques are not available. 

Instead, gravitational contributions are found by inserting factors of the stress tensor into matter correlation 
functions. This is because the graviton $h_{\mu \nu}$ couples to the stress tensor, as seen by expanding the 
matter action around the classical AdS$_2$ solution,
$g_{\mu \nu} = g^{(c)}_{\mu 
\nu} + h_{\mu \nu}$, giving 
\be \label{eq:GravAction}
S_{M}[g] = S_{M}[g^{(c)}] + \int \frac{\delta S}{\delta g_{\mu \nu}} h_{\mu \nu} 
+ \frac{1}{2} \int h_{\alpha \beta} \frac{\delta^2 S}{\delta g_{\alpha \beta} \delta g_{\mu \nu}} h_{\mu \nu}
+ O(h^3) \; ,
\ee
where the stress tensor $T^{\mu \nu}(x) = (2/\sqrt{g}) \delta S_{M} / \delta g_{\mu \nu}(x)$.  
Gravitational correlators are then found by computing the path integral
\be \label{eq:GravCorrDef}
\< \CO \>_{\text{grav}} = \int [Dh] \,  \< \CO \> \, e^{-S_{M}[h]} \; ,
\ee
where $\< \CO \>$ is the matter correlator. 
Expanding the action (\ref{eq:GravAction}) and the exponential in (\ref{eq:GravCorrDef}) generates $1/c$ 
gravitational corrections.

For example, suppose we want to calculate the leading $1/c$ gravitational correction to a boundary 
scalar two-point function $\< \tilde{\phi} \tilde{\phi} \>$ where $\tilde{\phi}$ is the CFT$_1$ dual to $\phi$. 
The correction comes from expanding (\ref{eq:GravAction}) in the path integral, resulting in a term
\be \label{eq:1/ccorrection}
\int d^2 x \, d^2 y \sqrt{g_x} \sqrt{g_y} \, \< \phi \phi T(x) T(y) \> \, H(x,y) \; ,
\ee
where $H(x,y) = \<h(x) h(y) \>$ is the graviton propagator. Doing this to completion requires calculating $ 
\< \phi \phi T(x) T(y) \>$, working out the $hT$ coupling in the bulk as well as the form of $H(x,y)$, and then 
integrating the above directly before taking the result to the boundary.\footnote{We are being schematic 
because we will not calculate gravitational corrections this way. A similar calculation is detailed in 
\cite{Anand:2017dav}.}
 
More generally, being able to compute gravitational corrections like (\ref{eq:1/ccorrection}) requires 
knowing the $TT$ OPE. For a general non-Gaussian CFT we do not know this as for $d>2$ the $TT$ OPE 
does not have a universal structure \cite{Osborn:1993cr}. Conformal invariance via the Ward identities 
\cite{Belavin:1984vu} shows how to insert factors of $T$ into $n$-point matter correlators:
\begin{align*} \label{eq:WardStressIdentity}
\< T(\xi)T(x_1)& \dots T(x_M) \phi_{1}(z_1) \dots \phi_{N}(z_N) \rangle \\
&= \left \{ \sum_{i=1}^{N} \left[ \frac{h_i}{(\xi - z_i)^2} + \frac{1}{\xi - z_i} \frac{\partial}{\partial z_i} \right] + 
\sum_{j=1}^{M} \left[ \frac{2}{(\xi - x_j)^2} + \frac{1}{\xi - x_j} \frac{\partial}{\partial x_j} \right] \right \} \\
&+ \sum_{j=1}^{M} \left[ \frac{c/2}{(\xi - x_j)^4} \right] \langle T(x_1) \dots T(x_{j-1}) T(x_{j+1}) \dots T(x_M) 
\phi_{1}(z_1) \dots \phi_{N}(z_N) \> \; . \numberthis
\end{align*}
Therefore the power of conformal invariance in two dimensions allows contributions from weakly-coupled 
gravity to be calculated, even if the matter sector is strongly coupled. 

However we are not going to calculate gravitational corrections for correlators in the 1d boundary theory 
this way; the same corrections can be streamlined using the 1d Schwarzian theory, as we will see below.

\subsection{Schwarzian effective action}

To calculate gravitational corrections using the Schwarzian theory 
we use Jackiw-Teitelboim (JT) gravity \cite{Jackiw:1984je,Teitelboim:1983ux} of the type studied in 
Refs.~\cite{Almheiri:2014cka, Maldacena:2016upp, Jensen:2016pah, Engelsoy:2016xyb}.
In Appendix \ref{app:JTSchwarzian} we review how the Schwarzian theory comes from JT gravity 
and how to calculate correlators in the Schwarzian theory, but it useful to summarize the salient ideas below.

JT gravity is a 2d dilaton gravity whose boundary description is given by a 1d Schwarzian action,
\be \label{eq:SchwarzAction}
S[t] = -C \int du \,  \{t, u \} \; , \quad \quad \{t, u \} \equiv \frac{t'''}{t'} - \frac{3}{2} \left( \frac{t''}{t'} \right)^2 \; ,
\ee
where $C$ is a bulk-valued coupling constant. We are  using the notation that the AdS$_2$ metric is given 
by $ds^2_{\text{AdS}_2} = z^{-2}(dt^2 + dz^2)$ and $u$ is the proper boundary time. Hence, $t(u)$ is the 
dynamical field variable and primes denote derivatives with respect to $u$. Because the Schwarzian action is just the 1d description of JT gravity, gravitational contributions are calculated via Schwarzian correlators,
\be
\< \tilde{\CO} \>_{t} = \int [dt] \, \< \tilde{\CO} \> \, e^{-S[t]} \; ,
\ee
where $\< \tilde{\CO} \>$ is the matter boundary correlator. 

The 1d Schwarzian theory is a theory of time reparametrizations whose action has a global SL(2,$\mathbbm{R}$) 
invariance. The fields $t(u)$ are Nambu-Goldstone 
bosons generated from the spontaneous breaking of SL(2,$\mathbbm{R}$) symmetry by the AdS$_2$ solution in the JT 
gravity action (\ref{eq:JTaction}) and represent the zero modes involving large diffeomorphisms 
\footnote{Although technically different, for curves (1-manifolds) reparametrizations and diffeomorphisms are 
the same.} As discussed in Appendix \ref{app:JTSchwarzian}, it is necessary to explicitly break the SL(2,$\mathbbm{R}$) 
symmetry by fixing the boundary length to be large but finite in order to lift these zero modes to non-zero 
modes needed for a dynamic gravity theory. 

By varying the Schwarzian action it can be shown that one of the solutions to the field 
equations of motion is a constant Schwarzian derivative (cf.~eq.~(\ref{eq:VarySchwarzAction})).
Indeed, for $t(u) = u$ the Schwarzian derivative vanishes, corresponding to a saddle point in the path 
integral language. Considering small reparametrizations, $t(u) \rightarrow t(u) = \tan \frac{u + \epsilon 
k(t)}{2}$, an effective local fluctuation action can be derived (\ref{eq:EffSchwarzAction}),
\be \label{eq:EffSchwAction}
S_{\epsilon}[t] = \frac{C}{2} \int_{0}^{2\pi} du \, \left[ (k'')^2 -  (k')^2 \right] \; ,
\ee
from which the fluctuation propagator (\ref{eq:flucprop}) can be found. 

In 1d reparametrizations and conformal transformations are the same. Therefore to calculate the leading 
perturbative gravitational correction we simply apply a conformal transformation to the matter correlator, 
\be
\tilde{\CO}(u_1, \dots , u_n) =  (t'_1)^{\Delta_1/2} \dots (t'_n)^{\Delta_n/2} \tilde{\CO}(t_1(u_1), \dots , 
t_n(u_n)) \; ,
\ee
where $t'_i \equiv d t_i(u_i)/d u_i$, and then use the effective Schwarzian action (\ref{eq:EffSchwAction})
to calculate the gravitational contribution.

\subsection{Non-Gaussian $\tilde{\phi}_{2,1}$ boundary four-point functions}
Aside from the Ising model, minimal model $\tilde{\phi}_{2,1}$ four-point functions are non-Gaussian. A 
proof of this statement can be found in Appendix \ref{app:NonGaussApp}. The Ising model has the 
factorizable spin boundary four-point function
\be \label{eq:4ptsigcorboundary}
\langle \sigma \sigma \sigma \sigma \rangle_{\partial} = \< \epsilon_1 \epsilon_2 \epsilon_3 \epsilon_4 \> 
= \frac{1}{t_{12} t_{34}} + 
\frac{1}{t_{13} t_{24}} + \frac{1}{t_{14} t_{23}} \; ,
\ee
a reflection of the fact that the Ising model contains the theory of a free fermion. Factorized four-point 
functions are a trivial extension of the two-point function and can be found in \cite{Maldacena:2016upp}.
Mixed four-point functions of the type $\< \phi_{h_1} \phi_{h_1} \phi_{h_2} \phi_{h_2} \>$ also fall into this 
category since taking them to the boundary produces a factorizable four-point function,
\be
\< \phi_{h_1} \phi_{h_1} \phi_{h_2} \phi_{h_2} \>_{\partial} =
\< \tilde{\CO}_{h_1} \tilde{\CO}_{h_1} \> \< \tilde{\CO}_{h_2} \tilde{\CO}_{h_2} \>
 = \frac{1}{t^{2\tilde{\Delta}_{\tilde{\CO}_1}}_{12}} \frac{1}{t^{2\tilde{\Delta}_{\tilde{\CO}_2}}_{34}} \; ,
\ee
of the type discussed before.

However, a boundary unitary minimal model $\mathcal{M}(m)$ four-point function $\< \tilde{\phi}_{2,1} 
\tilde{\phi}_{2,1} \tilde{\phi}_{2,1} \tilde{\phi}_{2,1} \>$ is given by a linear combination of two hypergeometric functions, each multiplied by factors of the cross-ratio,
\begin{align*} \label{eq:MMidentical4pt}
G_{4}(x) = (x(1-x))^{-\Delta} \bigg( {}_2 F_{1}\big(\frac{1-2\Delta}{3}, -2\Delta, \frac{2-4\Delta}{3}; x \big) 
+ A_{m} \, x^{\frac{1+4\Delta}{3}} {}_2 F_{1}\big(\frac{1-2\Delta}{3}, \frac{2+2\Delta}{3}, 
\frac{4+4\Delta}{3}; 
x \big) \bigg) \; , 
\numberthis
\end{align*}
where $A_{m} = e^{\frac{2\pi i}{m+1}} \frac{ \Gamma(\frac{2}{m+1}) 
\Gamma(2-\frac{3}{m+1})}{2\Gamma(\frac{1}{m+1}) 
\Gamma(\frac{2m}{m+1})}$, $\tilde{\Delta} \equiv 
\tilde{\Delta}_{\tilde{\phi}_{2,1}}$, 
and $x = \frac{t_{12} t_{34}}{t_{13} t_{24}}$. 
Under a conformal transformation 
this becomes
\be \label{eq:MM4ptconftrans}
G_{4}(t_1, t_2, t_3, t_4) \longrightarrow (f'_1 f'_2 f'_3 f'_4)^{\tilde{\Delta}} G_4(f_1, f_2, f_3, f_4) \; .
\ee
The gravitational correlator is found by plugging $f(t) = \tan \frac{t + \epsilon k(t)}{2}$ and 
(\ref{eq:MMidentical4pt}) into (\ref{eq:MM4ptconftrans}), expanding to quadratic order in $\epsilon$, and 
then contracting using the $\epsilon$-propagator. In practice, the easiest way to perform the calculation is to 
break it up into pieces. Start with the conformal transformation factor
\be
(f'_1 f'_2  f'_3 f'_4 )^{\tilde{\Delta}} \rightarrow F_{0}(1 + \epsilon F_{1} + \epsilon^2 F_{2}) + O(\epsilon^3)  
\;, 
\ee
where 
\begin{align*}
F_{0} &= 1 \\
F_{1} &= \tilde{\Delta} \sum_{i=1}^4 (k_i \tan \frac{t_i}{2} + k'_i ) \\
F_{2} &= \frac{\tilde{\Delta}}{4} \sum_{i=1}^4 ( k^2_i \sec^2 \frac{t_i}{2} - 2k'^2_i) 
+ \frac{\tilde{\Delta}^2}{2} \big(\sum_{i=1}^4 (k_i \tan \frac{t_i}{2} + k'_i )\big)^2 \; . \numberthis
\end{align*}
Next note that the transformation for the cross-ratio goes as
\be
x = \frac{f_{12} f_{34}}{f_{13} f_{24}} \rightarrow x_0 + \epsilon x_1 + \epsilon^2 x_2 + O(\epsilon^3) 
\ee
where 
\begin{align*} \label{eq:crossratiocoeffs}
x_{0} &= \frac{ \sin \frac{t_{12}}{2} \sin\frac{t_{34}}{2} }{\sin \frac{t_{13}}{2} \sin\frac{t_{24}}{2}}    \\
x_{1} &= \frac{1}{8} \csc^2 \frac{t_{13}}{2} \csc^2 \frac{t_{24}}{2} 
 \left( k_{12} \sin t_{34} + k_{31} \sin t_{24} +  k_{14} \sin t_{23} +
  k_{23} \sin t_{14} +  k_{42} \sin t_{13} + k_{34} \sin t_{12}   \right) \\
  x_{2} &= 
\left[ -\frac{k^2_1}{8} \csc^4 \frac{t_{13}}{2} \csc \frac{t_{24}}{2} \sin \frac{t_{23}}{2} \sin \frac{t_{34}}{2} \sin 
t_{13} 
+ \left( \begin{matrix} 1 \leftrightarrow 2 \\ 3 \leftrightarrow 4 \end{matrix} \right)
+ \left( \begin{matrix} 1 \leftrightarrow 3 \\ 2 \leftrightarrow 4 \end{matrix} \right)
+ \left( \begin{matrix} 1 \leftrightarrow 4 \\ 2 \leftrightarrow 3 \end{matrix} \right) \right] \\
&+ \left[ \frac{ k_1 k_2 }{4}  \csc^2 \frac{t_{13}}{2}  \csc^2 \frac{t_{24}}{2} \csc^2 \frac{t_{34}}{2} 
+ \left( \begin{matrix} 1 \leftrightarrow 3 \\ 2 \leftrightarrow 4 \end{matrix} \right)         \right] \\
&- \bigg[ \bigg( \frac{k_1 k_4}{4} \csc^2 \frac{t_{13}}{2}  \csc^2 \frac{t_{23}}{2} \csc^2 \frac{t_{24}}{2} \\
&+ \frac{k_1 k_3}{8} \left( \cos \frac{t_{13} + t_{24}}{2} - 2 \cos \frac{t_{12} + t_{34}}{2} + \cos \frac{t_{13} 
+ t_{42}}{2} \right)
 \csc^3 \frac{t_{13}}{2} \csc \frac{t_{24}}{2} \bigg)
+ \left( \begin{matrix} 1 \leftrightarrow 2 \\ 3 \leftrightarrow 4 \end{matrix} \right)  \bigg] \; . \numberthis
\end{align*}
Expanding (\ref{eq:MMidentical4pt}) results in
\be \label{eq:FourGrav}
G_{4}(x) = \mathcal{G}_{0} + \epsilon^2 \mathcal{G}_2
\ee
with
\begin{align*} 
\mathcal{G}_0 &= F_0 (x_0(1-x_0 ))^{-\tilde{\Delta}} 
\left[ {}_2F_1\left( \frac{1-2\tilde{\Delta}}{3}, -2\tilde{\Delta}, \frac{2-4\tilde{\Delta}}{3}; x_0 \right) 
+ A_m x^{\frac{1+4\tilde{\Delta}}{3}}_0 {}_2F_1\left( \frac{1-2\tilde{\Delta}}{3}, \frac{2+2\tilde{\Delta}}{3}, 
\frac{4+4\tilde{\Delta}}{3}; x_0 \right) \right] \\
\mathcal{G}_2 &= 
\frac{1}{6}(x_0(1-x_0))^{-\tilde{\Delta}}
\bigg[ \frac{1}{(x_0(x_0-1))^2} 3 ( 2 F_2 (x_0(x_0-1))^2 + \tilde{\Delta}(-2 F_1 x_0 x_1 (1-3x_0 + 2x^2_0) 
\\
&-2 
F_0 x_0 x_2 (1-3 x_0 + 2 x^2_0)  + F_0 
x^2_1(1+\tilde{\Delta}-2x_0(1+2\tilde{\Delta})+x^2_0(2+4\tilde{\Delta}) 
) ) )  \\  
&(A_m x^{\frac{1+4\tilde{\Delta}}{3}}_0 {}_2F_1\left( \frac{1-2\tilde{\Delta}}{3}, \frac{2+2\tilde{\Delta}}{3}, 
\frac{4+4\tilde{\Delta}}{3}; x_0 \right) 
+ {}_2F_1\left( \frac{1-2\tilde{\Delta}}{3}, -2\tilde{\Delta}, \frac{2-4\tilde{\Delta}}{3}; x_0 \right) \\
&+ \frac{ x_1(F_1 x_0(1-x_0) + F_0 x_1 \tilde{\Delta} (2 x_0 -1))   }{(x_0-1)x^{5/3}_0}
\bigg(-2 A_m x_0^{4\tilde{\Delta}/3} (1+4\tilde{\Delta}) {}_2F_1\left( \frac{1-2\tilde{\Delta}}{3}, 
\frac{2+2\tilde{\Delta}}{3}, \frac{4+4\tilde{\Delta}}{3}; x_0 \right) \\
&+ 6 \tilde{\Delta} x_0^{2/3}  {}_2F_1\left( 1-2\tilde{\Delta}, \frac{4-2\tilde{\Delta}}{3}, 
\frac{5-4\tilde{\Delta}}{3}; 
x_0 \right)
+ A_m x_0^{\frac{1+4\tilde{\Delta}}{3}}(-1+2\tilde{\Delta}) {}_2F_1\left( \frac{4-2\tilde{\Delta}}{3}, 
\frac{5+2\tilde{\Delta}}{3}, \frac{7+4\tilde{\Delta}}{3}; x_0 \right) \bigg) \\
&+ 6 F_0 \bigg( -\tilde{\Delta} x_2 \, {}_2F_1\left( 1-2\tilde{\Delta}, \frac{4-2\tilde{\Delta}}{3}, 
\frac{5-4\tilde{\Delta}}{3}; x_0 \right) 
+ \frac{x^2_1 \tilde{\Delta}(\tilde{\Delta} -2)(1-2\tilde{\Delta})}{5-4\tilde{\Delta}} 
{}_2F_1\left( 
2-2\tilde{\Delta}, \frac{7-2\tilde{\Delta}}{3}, \frac{8-4\tilde{\Delta}}{3}; x_0 \right) \\
&+ \frac{A_m}{18} x^{\frac{-5+4\tilde{\Delta}}{3}} \bigg(2(1+4\tilde{\Delta})(3 x_0 x_2 
+x^2_1(-1+2\tilde{\Delta})) 
{}_2F_1\left( \frac{1-2\tilde{\Delta}}{3}, \frac{2+2\tilde{\Delta}}{3}, \frac{4+4\tilde{\Delta}}{3}; x_0 \right) \\
&+ \frac{ x^2_0 x^2_1(\tilde{\Delta} -2)(-1+2\tilde{\Delta})(5+2\tilde{\Delta})  }{7+4\tilde{\Delta}} 
{}_2F_1\left( 
\frac{7-2\tilde{\Delta}}{3}, \frac{8+2\tilde{\Delta}}{3}, \frac{10+4\tilde{\Delta}}{3}; x_0 \right) \\
&+ x_0(1-2\tilde{\Delta})(3 x_0 x_2 + x^2_1(1+4\tilde{\Delta})) {}_2F_1\left( \frac{4-2\tilde{\Delta}}{3}, 
\frac{5+2\tilde{\Delta}}{3}, \frac{4+4\tilde{\Delta}}{3}; x_0 \right) \bigg) \bigg) 
\bigg] \numberthis
\end{align*}
after expanding the hypergeometrics and $x(1-x)$ prefactor. The final expression is found by inserting 
(\ref{eq:crossratiocoeffs}) into the above and contracting using the fluctuation propagators (\ref{eq:flucprop}). 
The linear term vanishes due to $\< \epsilon \>$ terms, which can also serve as a calculational check. 

The point of the above is to explicitly show that we can in principle fully calculate to arbitrary order the 
gravitational contribution to unitary minimal model $\tilde{\phi}_{2,1}$ four-point functions, which aside 
from the Ising model are non-Gaussian. The generalization to higher $n$-point functions on the boundary is 
straight forward, if tedious.

\section{Higher Dimensions} \label{sec:HigherDims}

So far we have focused on two dimensions in the bulk, however, we can extend some of the previous 
discussion to the higher dimensional case of CFT$_{d+1}$ in AdS$_{d+1}$ with $d>1$. We aim to outline 
how to calculate gravitational corrections to the CFT$_d$ boundary correlators like for the $d=1$ case of 
Sec.~\ref{sec:GravCoupling}. 

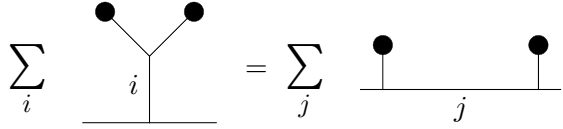
\begin{figure}
             \begin{center}           
              \includegraphics[scale=0.6]{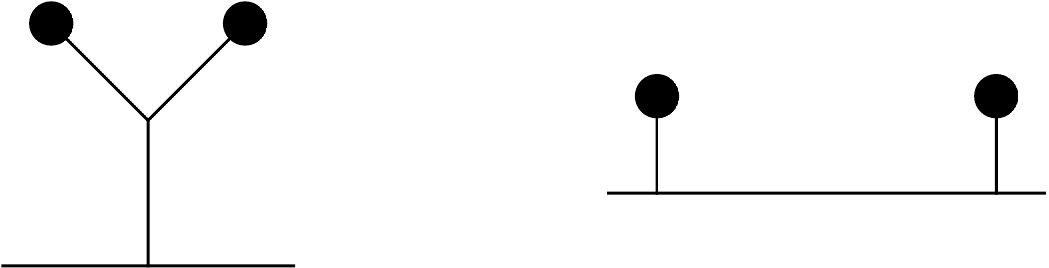}
              \put(-210,18){{\Large $\displaystyle \sum_i$}}              
              \put(-120,18){=}
              \put(-105,18){{\Large $\displaystyle \sum_j$}}
              \put(-164,11){$i$}
              \put(-41,3){$j$}                                    
              \caption{Two-point function crossing symmetry in BCFT$_d$.}
              \label{fig:BoundBoot}
            \end{center}
	\end{figure}

First, let us review a bulk scalar two-point function $\< \phi \phi \>$ in AdS$_{d+1}$ without gravity.
Similar to before, after a coordinate transformation and Weyl transformation this can be mapped to 
the flat UHP metric $ds^2 = y^{-2}(d\vec{x}^2 + dy^2)$ with $x^{\mu} = (y, \vec{x})$ and $\vec{x} \in 
\mathbbm{R}^d$. A two-point function in the presence of the $y=0$ boundary will obey a crossing symmetry 
equating the bulk and boundary decompositions. A thorough analysis of this bootstrap for boundary 
CFT$_d$ (BCFT$_d$) was done in \cite{Liendo:2012hy}. The statement of crossing symmetry for the 
two-point scalar function in BCFT$_d$ is given by
\be \label{eq:BoundBootEq}
1+ \sum_{\CO'} C_{\phi \phi \CO'} A_{\CO'} f_{\text{bulk}}(\Delta_{\CO'}; \xi) 
= \xi^{\Delta} \left( A_{\phi}^2 + \sum_{\tilde{\CO}} C^2_{\phi \tilde{\CO}}  
f_{\text{bdry}}(\Delta_{\tilde{\CO}}; 
\xi)  \right) \; ,
\ee
where $\xi = (x_1 - x_2)^2/(4 y_1 y_2)$ and $f_{\text{bulk}}$, $f_{\text{bdry}}$ are the respective bulk and 
boundary conformal blocks given by
\begin{align*} \label{eq:TwoPointBlocks}
f_{\text{bulk}}(\Delta; \xi) &= \xi^{\Delta/2} {}_2 F_1 
\left( \frac{\Delta}{2}, \frac{\Delta}{2}, \Delta + 1 -\frac{d}{2}; - \xi \right) \\
f_{\text{bdry}}(\Delta; \xi) &= \xi^{-\Delta} {}_2 F_1 
\left( \Delta, \Delta + 1 -\frac{d}{2}, 2\Delta + 2 - d; - \frac{1}{\xi} \right) \numberthis \; .
\end{align*}
Crossing (\ref{eq:BoundBootEq}) is pictorially encapsulated in Fig.~\ref{fig:BoundBoot}.
The power of this bootstrap approach is that given the bulk data we can solve for the boundary data, 
which is necessary for computing boundary gravitational correlators.

Therefore, suppose we are given the bulk data and thus can calculate the boundary data. Recall that an 
expansion in the AdS$_{d+1}$ gravitational constant $G$ corresponds to a $1/c$ expansion on the 
CFT$_d$ side. Therefore the leading correction to the two-point gravitational correlator is given by the 
gravitational loop diagram in the bottom-right of Fig.~\ref{fig:WeakVsStrong}. This corresponds to calculating 
the higher-dimensional counterpart to (\ref{eq:1/ccorrection}),  
\begin{align*} \label{eq:AdSBulkGravIntegral}
\int d^{d+1} x_1 d^{d+1} x_3 \sqrt{g_{x_1}} \sqrt{g_{x_3}}
\< T_{\mu \nu}(x_1) \tilde{\phi}(x_2) T_{\rho \sigma}(x_3) \tilde{\phi}(x_4) \> G^{\mu \nu, \rho \sigma}_{BB}(x_1, x_3) 
\; , \numberthis
\end{align*}
where $G^{\mu \nu, \rho \sigma}_{BB}$ is the $(d+1)$-dimensional graviton bulk propagator 
\cite{DHoker:1999bve}. On the CFT$_d$ side, this correponds to our object of interest, a $1/c$ expansion in 
the two-point boundary correlator 
\be \label{eq:1/cBoundBound}
\< \tilde{\phi}(\vec{x}) \tilde{\phi}(0) \> = \frac{1 + \frac{\tilde{N}^{(1)}_{\phi}}{c} + \cdots 
}{\vec{x}^{2\tilde{\Delta}_{\phi}}} 
\left(1 - \frac{2 \tilde{\Delta}^{(1)}_{\phi}}{c} \ln \vec{x} 
- \frac{2 \tilde{\Delta}^{(2)}_{\phi}}{c^2} \ln \vec{x} + \frac{2 (\tilde{\Delta}^{(1)}_{\phi})^2}{c^2} \ln^2 \vec{x} 
+
\cdots \right)
\ee
after expanding the scaling dimension and normalization in $1/c$
\be
\tilde{\Delta} = \sum_{n=0}^{\infty} \frac{\tilde{\Delta}^{(n)}_{\phi}}{c^n} \; , \quad \quad
\tilde{N}_{\phi} = \sum_{n=0}^{\infty} \frac{\tilde{N}^{(n)}_{\phi}}{c^n} \; ,
\ee
due to the presence of a finite gravitational constant. 

To calculate (\ref{eq:AdSBulkGravIntegral}) it is convenient to first expand $T$ in terms of 
boundary quantities to be able to then work in a CFT$_d$ without a boundary. 
The bulk stress tensor can be expanded in boundary operators $\tilde{\CO}$ via
\be \label{eq:TBoundExpansion}
T^{\mu \nu}(y, \vec{x}) = \sum_{i} (2y)^{\tilde{\Delta}_i - d}  b^{\mu \nu}_{i, \alpha_1 \dots \alpha_l} \tilde{\CO}^{\alpha_1 \dots 
\alpha_l}_{i}(\vec{x}) \; , 
\ee 
corresponding to the configuration 
\begin{figure}[H]
\centering
\includegraphics[scale=.32]{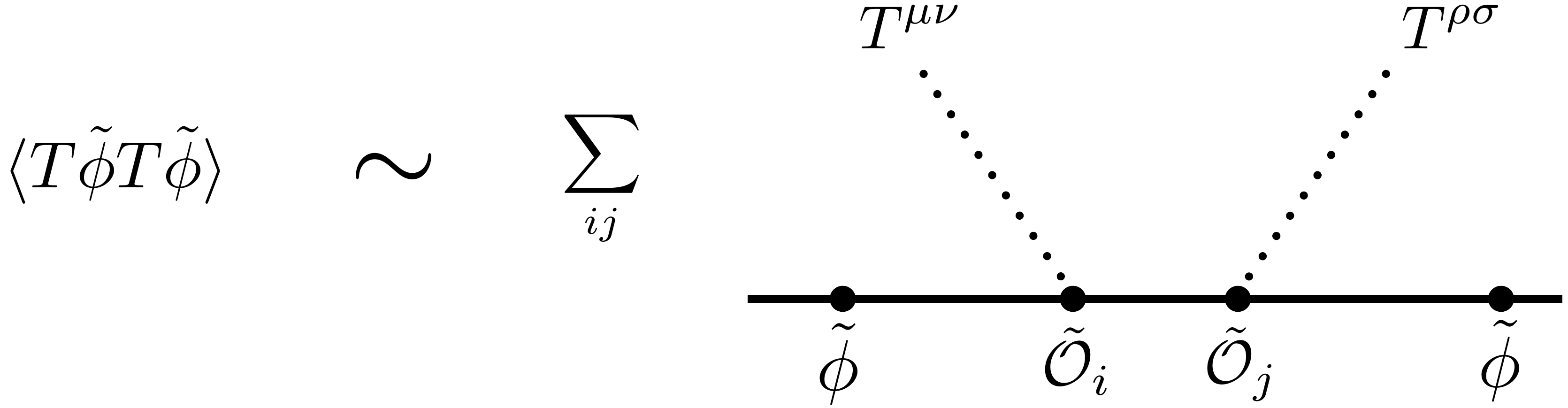}
\end{figure}
\vspace{-0.8cm}
\hspace{-0.88cm}
which only involves boundary operators.
Since all operators are on the boundary, a CFT$_d$, a 
conformal block decomposition for $\< \tilde{\phi} \tilde{\CO} \tilde{\phi} \tilde{\CO} \>$ applies:
\begin{figure}[H]
\centering
\includegraphics[scale=.33]{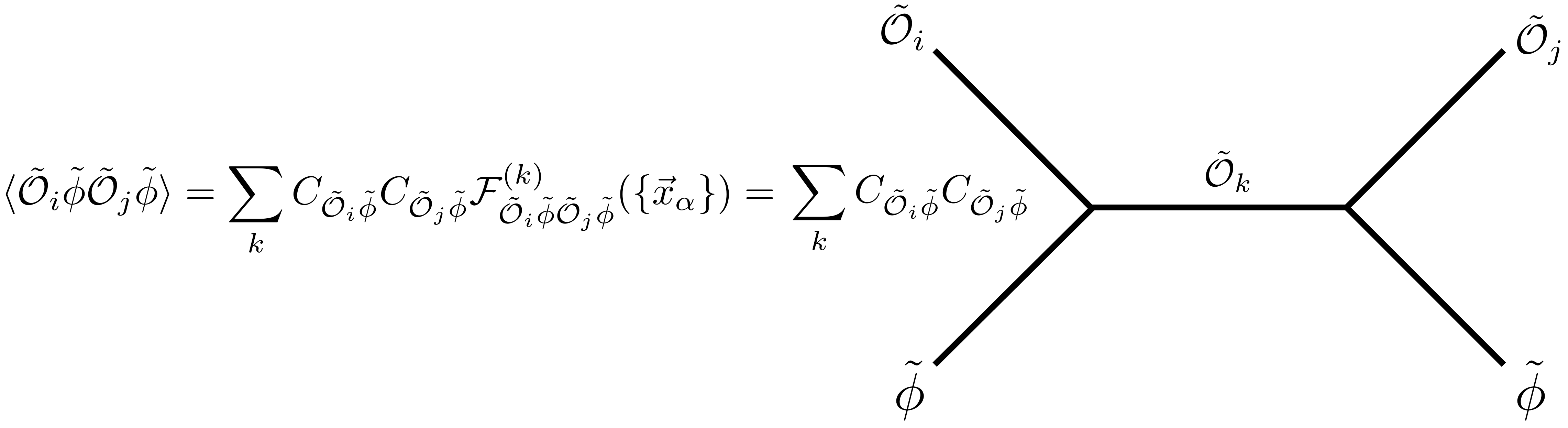}
\end{figure}
\vspace{-0.85cm}
\hspace{-0.88cm}
Putting it all together gives 
\begin{align*} \label{eq:TpTp}
\< T(x_1) \tilde{\phi}(x_2) T(x_3) \tilde{\phi}(x_4) \> = \sum_{ijk} (4 y_2 y_4)^{\tilde{\Delta}_{\phi} - \Delta_{\phi}}
(2y_1)^{\tilde{\Delta}_i - d}
(2y_3)^{\tilde{\Delta}_j - d}
C^2_{\phi \tilde{\phi}} b^{\mu \nu}_i b^{\rho \sigma}_j
C_{\tilde{\CO}_i \tilde{\phi}} C_{\tilde{\CO}_j \tilde{\phi}} 
\mathcal{F}^{(k)}_{\tilde{\CO}_i \tilde{\phi}\tilde{\CO}_j \tilde{\phi}}(\{\vec{x}_{\alpha} \}) \; . \numberthis
\end{align*}
for the bulk stress tensor-scalar four-point function. After inserting (\ref{eq:TpTp}) into (\ref{eq:AdSBulkGravIntegral}) and calculating the integrals, the result is the leading gravitational correction to the CFT$_d$ boundary scalar two-point function: the $O(1/c)$ term in (\ref{eq:1/cBoundBound}).

\section{Discussion} \label{sec:Conclusions}
Analyzing minimal models in AdS$_2$ allows us to take a strongly coupled CFT and use the power of 
boundary CFT and JT/Schwarzian gravity. The result  is a family of ``solved" bulk theories that we can 
calculate perturbative gravitational corrections for to arbitrary order. We illustrated examples of the types of 
calculations that can be done in this framework using the Ising model, which can be generalized to any 
minimal model. In particular we saw how the physical boundary conditions affect the operator content of the 
theory through the relation between the boundary conditions, correlator coefficients, and the OPE 
coefficients. For the case of AdS with two boundaries -- as found, e.g., in RS1 models -- we calculated the energy 
spectrum from the boundary correlators. JT gravity was added and the leading gravitational correction to the 
non-Gaussian $\tilde{\phi}_{2,1}$ minimal model correlators was found.

It is interesting to think about the scenario studied here to the arguments put forth in 
\cite{Heemskerk:2009pn}. In that work, it is conjectured that a CFT with a large-$N$ expansion and a 
sufficiently large gap has a local bulk dual description. In the case at hand, the bulk is strongly coupled and 
the boundary operator dimensions and correlators do not have a large-$N$ factorization. Therefore, in a 
sense holography is more general than the requirements of \cite{Heemskerk:2009pn}.

\section*{Acknowledgements}
This work grew out of a collaboration with Liam Fitzpatrick, to whom I owe many thanks for his insight 
and guidance. We also thank Hongbin Chen for helpful discussions.
This work was supported in part by the Simons Collaboration on the Nonperturbative Bootstrap and Boston University.

\appendix

\section{OPE Conventions} \label{app:conventions}
In this work we deal with bulk, boundary, and bulk-boundary operator product expansions, with the 
additional complication of these being dependent on potentially changing boundary conditions. Here we 
define the OPE notations and conventions used throughout the text, based on the notation used in 
\cite{Lewellen:1991tb}. 

For the two-dimensional bulk theory, conformal invariance fixes the form of the one- and two-point 
functions. We normalize the vacuum and the primary field $\CO_i$ with scaling dimensions $(\Delta_i,\bar{\Delta}_i)$ as
\be
\< \mathbbm{1} \> = 1, \quad \<\CO_i \> &= 0 \; , \quad \< \CO_i(z,\bar{z}) \CO_j(0) \> = \delta_{ij} \, z^{-2 
\Delta_i} \bar{z}^{-2 
\bar{\Delta}_i} \, .
\ee
At short distances the bulk OPE is given by
\be
\CO_i(z,\bar{z}) \CO_j(0) = \sum_k C_{ijk} \, z^{\Delta_k - \Delta_i - \Delta_j} \bar{z}^{\bar{\Delta}_k - 
\bar{\Delta}_i - \bar{\Delta}_j} 
\CO_k(z,\bar{z}) + \cdots
\ee
Restricting ourselves to the one-dimensional boundary, we now have operators $\COt_i^{ab}(x)$ that are 
primaries under a single Virasoro algebra with conformal dimension $\tilde{\Delta}_i$. The superscript 
indices $ab$ indicate a boundary operator that mediates a change in boundary conditions from type $a$ to 
type $b$. The boundary correlators are thus
\be  \label{eq:boundtwopoint}
\< \mathbbm{1} \>_a = \alpha^a, \quad \<\COt_i^{ab} \> &= 0 \; , \quad \< \COt_i^{ab}(0) \COt_j^{cd}(x) \> 
= 
\delta_{ad}\delta_{bc}\delta_{ij} 
\, x^{-2 \tilde{\Delta}_i} \tilde{\alpha}^{ad}_i \; ,
\ee
with a boundary OPE given by
\be \label{eq:boundope}
 \COt_i^{ab}(0) \COt_j^{cd}(x) = \sum_k \tilde{C}^{abc}_{ijk} \, x^{\tilde{\Delta}_k - \tilde{\Delta}_i - 
\tilde{\Delta}_j} \COt_k^{ac}(x) + \cdots
\ee
The normalization constants $\alpha^a$ and $\alpha^{ad}_i$ are boundary condition-dependent and arise 
from that fact that having canonically fixed the bulk normalizations, we can not do the same with the 
boundary ones. 

When a bulk operator approaches a boundary with boundary condition $a$, we get a bulk-boundary OPE
\be \label{eq:bulkboundOPE}
\CO_j(z,\bar{z}) = \sum_i (2 y)^{\tilde{\Delta}_i - \Delta_j - \bar{\Delta}_j} C_{ji}^a \, \COt_i^{aa}(x) + \cdots
\ee
Due to the single Virasoro algebra, the explicit factor of 2 comes from thinking about 
(\ref{eq:bulkboundOPE}) as one chiral half of the 
bulk OPE.

For the sake of notational easement we drop the boundary condition superscripts when dealing with 
constant and uniform 
boundary conditions or otherwise when it is unambiguously clear (e.g., $\COt_i^{aa} \rightarrow \COt_i$, 
$C_{ji}^a \rightarrow C_{ji}$).

\section{The Modular Matrix $S$} \label{app:ModMatrix}
The purpose of this appendix is to show the relationship between the modular matrix $S$ and the fusion 
matrices and algebra explicitly for the Ising model. The matrix elements of $S$ are given by
\be \label{eq:Selements}
S_{rs;\rho \sigma} = 2 \sqrt{    \frac{2}{pq} } (-1)^{1+s\rho + r\sigma} \sin(\pi \frac{p}{q}r \rho) \sin(\pi 
\frac{q}{p}s \sigma) \, .
\ee 
For the minimal models, the central charge can be written as
\be
c = 1 - 6 \frac{(p-q)^2}{pq} \; ,
\ee
which corresponds to the irreducible representation with Kac indices $(r,s)$ in the range
\begin{align*} \label{eq:MMindicesrange}
1 \leq r& \leq q - 1 \\
1 \leq s& \leq p - 1 \\
qs &< pr \, . \numberthis
\end{align*}
It is common to denote $E_{p,q}$ as the set of pairs $(r,s)$ in the above range (\ref{eq:MMindicesrange}) 
and that $\rho, \sigma \in E_{p,q}$ in (\ref{eq:Selements}). This means that if $(q,p) = (4,3)$ corresponding to the Ising model, the allowed pairs are $(r,s): \{ (1,1), \, (2,1), \, (2,2) \}$. Indeed, plugging the allowed pairs into (\ref{eq:Selements}) we obtain the following matrix elements:
\begin{align*}
S_{11,11} = \frac{1}{2} \; , \quad S_{11,21} &= \frac{1}{2} \; , \quad  S_{11,22} = \frac{1}{\sqrt{2}} \; , \\
S_{21,11} = \frac{1}{2} \; , \quad S_{21,21} &= \frac{1}{2} \; , \quad  S_{21,22} = -\frac{1}{\sqrt{2}} \; , \\
S_{22,11} = \sqrt{\frac{1}{2}} \; , \quad S_{22,21} &= -\sqrt{\frac{1}{2}} \; , \quad  S_{22,22} = 0 \; , 
\numberthis
\end{align*}
corresponding to the Ising modular matrix $S$:
\be \label{eq:IsingS}
S_{\text{Ising}} = 
\begin{pmatrix}
\frac{1}{2} & \frac{1}{2}  & \sqrt{\frac{1}{2}} \\
\frac{1}{2} & \frac{1}{2}  & -\sqrt{\frac{1}{2}} \\
\sqrt{\frac{1}{2}} & -\sqrt{\frac{1}{2}}  & 0
\end{pmatrix} \; .
\ee
Recall that $S$ is a transformation on the basis of the minimal characters,
\be
\chi_{r,s}(-1/\tau) = \sum_{(\rho,\sigma) \in E_{p,q}} S_{rs, \rho \sigma} \chi_{\rho, \sigma}(\tau) \; ,
\ee
and can be calculated from explicit values for the characters $\chi$. In the Ising case: $\chi_{0}(\tau), \, 
\chi_{\frac{1}{2}}(\tau), \,$ and $\chi_{\frac{1}{16}}(\tau),$ respectively. The characters involve theta functions and Dedekind eta functions. The fact that under a modular transformation $S: \tau \mapsto -1/\tau$
\be
\chi_{\frac{1}{16}}(\tau) \mapsto \chi_{\frac{1}{16}}(-1/\tau) = \frac{1}{\sqrt{2}} \chi_{0}(\tau) - 
\frac{1}{\sqrt{2}} 
\chi_{\frac{1}{2}}(\tau) + 0 
\cdot \chi_{\frac{1}{16}}(\tau)
\ee
is reflected in the vanishing matrix element in the Ising matrix $S$. 

The fusion matrices can be calculated from the Verlinde formula,
\be
N^{k}_{\; ij} = \sum_{l=0}^2 \frac{ S_{jl} S_{il} (S^{-1})_{lk}}{S_{0l}} \; ,
\ee
to be
\be \label{eq:Ising_fusion}
N_{\mathbbm{1}} = 
\begin{pmatrix}
1 & 0 & 0 \\
0 & 1 & 0 \\
0 & 0 & 1
\end{pmatrix} \; , \quad
N_{\epsilon} = 
\begin{pmatrix}
0 & 1 & 0 \\
1 & 0 & 0 \\
0 & 0 & 1
\end{pmatrix} \; , \quad
N_{\sigma} = 
\begin{pmatrix}
0 & 0 & 1 \\
0 & 0 & 1 \\
1 & 1 & 0
\end{pmatrix} \; ,
\ee
so that 
\be
S \; N_{\mathbbm{1}} \; S = \text{diag}(1,1,1) \; , \quad S \; N_{\epsilon} \; S = \text{diag}(1,1,-1) \; , \quad 
S \; N_{\sigma} \; S = 
\text{diag}(\sqrt{2},-\sqrt{2},0) \; .
\ee
From (\ref{eq:Ising_fusion}) and the fusion algebra $[\Phi_i] \times [\Phi_j] = \sum_k N^{k}_{\; ij} [\Phi_k] $ 
we can read off the fusion rules 
\begin{align*}
\sigma \times \sigma &= \mathbbm{1} + \epsilon \\
\sigma \times \epsilon &= \sigma \\
\epsilon \times \epsilon &= \mathbbm{1} \; . \numberthis
\end{align*}

\section{Boundary States and Bulk-Boundary Constants}

Cardy boundary states are the physical boundary states and can be represented as a linear combination 
of boundary states that preserve local conformal invariance, so-called Ishibashi states. Given $S$, a Cardy 
boundary state $|\tilde{l}\>$ can be expanded in the basis of Ishibashi states $|j\rrangle$ corresponding to 
the bulk primary fields as \cite{Cardy:1989ir}
\be \label{eq:Sbasisexpansion}
| \tilde{l} \> = \sum_{j} \frac{S_{ij}}{\sqrt{S_{0j}}} |j \rrangle \; .
\ee
The matrix $S$ also encodes the coupling between bulk fields and the identity on the boundary through 
\cite{CARDY1991274}
\be \label{eq:onepointboundcoupling}
C^{a}_{j \mathbbm{1}} = A^{a}_j = \frac{ \llangle j | a \> }{ \llangle 0 | a \>} = \frac{S_{ja}}{S_{0a}} 
\left(\frac{S_{00}}{S_{j0}}\right)^{1/2} \; ,
\ee
where $j$ is an Ishibashi state and $a$ is a Cardy boundary state defined in (\ref{eq:Sbasisexpansion}). 
Applying the above to the Ising modular matrix (\ref{eq:IsingS}), the Cardy boundary states are given by 
\begin{align*} \label{eq:IsingBoundaryStates}
|\tilde{0} \> &= \frac{1}{\sqrt{2}} |0 \rrangle + \frac{1}{\sqrt{2}} |\epsilon \rrangle + \frac{1}{2^{1/4}} |\sigma 
\rrangle \\
|\tilde{\frac{1}{2}} \> &= \frac{1}{\sqrt{2}} |0 \rrangle + \frac{1}{\sqrt{2}} |\epsilon \rrangle - \frac{1}{2^{1/4}} 
|\sigma 
\rrangle \\
|\tilde{\frac{1}{16}} \> &= |0 \rrangle - |\epsilon \rrangle \; . \numberthis
\end{align*}
corresponding to the all spins pointing up $(+)$, down $(-)$, or left free $(f)$, respectively. The boundary 
operators corresponding to these changes 
in boundary conditions are $\tilde{\epsilon}^{\pm \mp}(x)$, $\tilde{\epsilon}^{ff}(x)$, $\tilde{\sigma}^{\pm 
f}$, and $\tilde{\sigma}^{f \pm}(x)$, with $\Delta_{\tilde{\epsilon}} = 1/2$ and $\Delta_{\tilde{\sigma}} = 1/16$. 
These can be calculated from the fusion coefficients $n^{h}_{\alpha \beta} = N^{h}_{\alpha \beta}$ and are 
derived in \cite{CARDY1986200}.

The modular matrix can also be used to calculate the bulk-boundary coefficients. Directly applying 
(\ref{eq:onepointboundcoupling}) 
to (\ref{eq:IsingS}) gives
\be \label{eq:bulkboundconst1}
C^{\pm}_{\tilde{\sigma} \mathbbm{1}} = A^{\pm}_{\sigma} = \pm 2^{1/4}  \; , \quad \quad 
C^{f}_{\tilde{\sigma} 
\mathbbm{1}} = 
A^{f}_{\sigma} 
= 0 \; ,
\ee
\be \label{eq:bulkboundconst2}
C^{\pm}_{\tilde{\epsilon} \mathbbm{1}} = A^{\pm}_{\epsilon} = 1  \; , \quad \quad C^{f}_{\tilde{\epsilon} 
\mathbbm{1}} = A^{f}_{\epsilon} 
= -1 \; ,
\ee
\be \label{eq:bulkboundconst4}
\left( C^{f}_{\tilde{\sigma} \tilde{\epsilon}}   \right)^2 = \frac{1}{\sqrt{2} \alpha^{ff}_{\tilde{\epsilon}}} \; ,
\ee 
for the bulk-boundary (and half-plane bulk one-point) structure constants. These are consistent with the 
symmetries of the Ising model under spin reversal $(\sigma \rightarrow - \sigma, \; \epsilon \rightarrow 
\epsilon, \; |+\> \leftrightarrow |-\>, \; |f\> \rightarrow |f\>)$ and duality $(\epsilon \rightarrow - \epsilon, \; |\pm 
\> \leftrightarrow |f \>)$.

\section{Non-Gaussianity of Minimal Model $\phi_{2,1}$ Four-Point Functions} \label{app:NonGaussApp}

Apart from the Ising model, unitary minimal model $\phi_{2,1}$ four-point functions are non-Gaussian.  
A ``Gaussian" correlator is one that is entirely determined by its two-point functions. This means that the 
connected four-point function of a Gaussian correlator tends to zero when operator pairs are pulled apart 
from each 
other due to clustering. The factorization of the Ising energy correlator (\ref{eq:4ptepscor}) is an example of 
this, which can be seen equivalently through
\be
\lim_{\stackrel{x_2 \rightarrow x_1}{x_4 \rightarrow x_3}} \frac{\<\epsilon_1 \epsilon_2 \epsilon_3 
\epsilon_4 \>}{\<\epsilon_1 \epsilon_2 \> \<\epsilon_3 \epsilon_4 \>} \rightarrow 1 \; .
\ee
Therefore, a test that a generic four-point correlator of the operator $\CO$ is Gaussian is that
\be \label{eq:GaussianTest}
\lim_{\stackrel{x_2 \rightarrow x_1}{x_4 \rightarrow x_3}} x_{12}^{2\Delta_{\CO}} x_{34}^{2\Delta_{\CO}} 
G_{4}^{\CO}(x_i) \rightarrow 1 \; ,
\ee
where $G_{4}^{\CO}(x_i)$ is given by
\be \label{eq:Full4pt}
G_{4}^{\CO}(x) = \frac{1}{\left|x(1-x)\right|^{2\Delta}} \left( \left|\mathcal{F}^{\CO}_1(x)\right|^2 + A^2_{m} 
\left|x\right|^{\frac{2+8\Delta}{3}} |\mathcal{F}^{\CO}_2(x)|^2 \right) \; .
\ee
with
\begin{align*}
\mathcal{F}^{\CO}_1(x) = {}_2 F_1 \left( \frac{1-2\Delta_{\CO}}{3}, -2\Delta_{\CO}, 
\frac{2-4\Delta_{\CO}}{3}; 
x \right) \; , \quad 
\mathcal{F}^{\CO}_2(x) = {}_2 F_1 \left( \frac{1-2\Delta_{\CO}}{3}, \frac{2+2\Delta_{\CO}}{3}, 
\frac{4+4\Delta_{\CO}}{3}; x \right) \; .
\end{align*}
and $|A_{m}|^2 = \Gamma(\frac{2}{m+1}) \Gamma(2-\frac{3}{m+1})/\Gamma(\frac{1}{m+1}) 
\Gamma(\frac{2m}{m+1})$.

It can be checked using (\ref{eq:GaussianTest}) that the first term in (\ref{eq:Full4pt}) is Gaussian for an 
operator of any scaling dimension whereas the second term is not. Therefore, to be fully Gaussian a 
four-point function needs the second term to vanish. From the fusion algebra
\be
\left[ \phi_{2,1} \right] \times \left[ \phi_{2,1} \right] = \left[ \phi_{1,1} \right] + \cdots 
\ee
this occurs when the higher operators denoted by the ellipses are absent in the Kac table of the minimal 
model in question associated with the $\phi_{2,1}$ operator. This only occurs for the Ising minimal model 
where $\phi_{2,1} = \epsilon$ and
\be
\left[ \epsilon \right] \times \left[ \epsilon \right] = \left[ \mathbbm{1} \right] \; .
\ee

\section{JT Gravity and the Schwarzian Action} \label{app:JTSchwarzian}
This section aims to concisely review how the Schwarzian action arises as the one-dimensional boundary 
theory to JT gravity, and then develops the effective Schwarzian theory. There are several good review articles \cite{Sarosi_2018, Trunin_2021} that cover much of this material in greater depth.

\subsection{JT gravity}
JT gravity is a theory with a two-dimensional metric tensor $g_{\mu \nu}$ and dilaton field $\Phi$ coupling 
given by the Euclidean action
\be \label{eq:JTaction}
S_{\text{JT}} = -\frac{1}{16\pi G} \left[ \int_{\mathcal{M}} d^2x \sqrt{g} \, \Phi \left(R - \frac{\Lambda}{L^2} 
\right)  + 2 
\int_{\partial\mathcal{M}}du \sqrt{h} \, \Phi_{b} K \right] \; ,
\ee
where $\mathcal{M}$ is the manifold the theory is defined on and $\partial \mathcal{M}$ is its boundary 
with induced metric $h$ and extrinsic curvature $K$.\footnote{We have suppressed the purely topological 
piece
\be
S_{\text{JT}} \supset \Phi_b \left( \int d^2 x \sqrt{g} R + 2 \int du \sqrt{h} K \right)
\ee
responsible for the extremal entropy.}
The bulk term uses the dilaton as a Lagrange multiplier to set the curvature to the cosmological constant $R 
= \Lambda/L^2 = -2$ in the equations of motion. The boundary term is the Gibbons-Hawking-York (GHY) 
term, which ensures a well-defined variational principle for the Dirichlet problem.  

To understand the boundary degrees of freedom the boundary proper length needs to be fixed.
This is done by cutting off the exactly Euclidean AdS$_2$ space to a nearly AdS$_2$ (NAdS$_2$) 
space with a bulk IR/boundary UV regulator by setting 
\be
\Phi_{b} = \Phi_{\text{bdy}} = \frac{\Phi_{r}(u)}{\epsilon} \; , \quad
\frac{1}{\epsilon^2} = g_{uu} = \frac{t'^2 + z'^2}{z^2}  \; , \quad
ds|_{\text{bdy}} = \frac{du}{\epsilon} \implies  \int_{0}^{\beta} ds = \frac{\beta}{\epsilon}
\ee
so that as $\epsilon \rightarrow 0$ the boundary proper length becomes large. The quantity $\Phi_r(u)$ 
is a dimension $-1$ coupling that is ``renormalized" in the sense it remains finite in the above limit. The 
primes are derivatives with respect to the proper boundary time $u$.

The result of this is a trajectory $(t(u), z(u))$ near the boundary corresponding to a UV cutoff in 
the dual boundary theory, producing a nearly CFT$_1$ (NCFT$_1$). Notice the parametrization $z(u) = 
\epsilon t'(u)$ satisfies the above equations at linear order in epsilon for arbitrary $t(u)$, so there is a family 
of boundary curves related by a reparametrization of $t(u)$. 

With this parametrization the induced metric $\sqrt{h} = 1/\epsilon$ and the remaining boundary term  
becomes
\be
S_{\text{JT}} \rightarrow -\frac{1}{8\pi G} \int \frac{du}{\epsilon} \frac{\Phi_{r}(u)}{\epsilon} K \; ,
\ee
where the extrinsic curvature is given by
\be
K = 1 + \epsilon^2 \{t, u \} + \cdots \quad  \quad  \{t, u \} \equiv \frac{t'''}{t'} - \frac{3}{2} \left( \frac{t''}{t'} 
\right)^2 \; ,
\ee
with $ \{t, u \}$ the Schwarzian derivative.
The result of this is the linearized boundary action of (\ref{eq:JTaction}) reduces to an effective 
Schwarzian action
\be \label{eq:SchwarzAction}
S = - \, C \int dt \, \Phi_{r}(u) \{t, u \}  \; ,
\ee
where $C=\frac{1}{8\pi G}$. The normalizable part of the dilaton, $\Phi_{r}(u)$, is constant on the 
boundary and serves as an external coupling.\footnote{On the boundary, $\Phi_{r}/ \epsilon = \Phi_b$, which 
can be absorbed into the definition of $C$.}
 
In the Schwarzian action (\ref{eq:SchwarzAction}), the dynamical field variable $t(u)$ represents the 
Nambu-Goldstone bosons generated from the spontaneous breaking of SL$(2)$ symmetry 
by the AdS$_2$ solution, and are zero modes involving large diffeomorphisms. In the conformal limit 
where $\beta = \infty \leftrightarrow T=0$, these zero modes have zero energy. Moving to NCFT$_1$ lifts these zero modes to nearly-zero modes needed for a dynamic gravity theory.

\subsection{Effective Schwarzian action}
By varying the action we get the equation of motion 
\be \label{eq:VarySchwarzAction}
\left[ \frac{1}{t'} \left( \frac{(t' \Phi_{r})'}{t'} \right) \right]' \longrightarrow \frac{ \{t, u \}'  }{t'} = 0
\ee
where in moving from left to right we used that $\Phi_r$ is constant on the boundary. We see that when $t$ 
is 
a linear function of $u$, the equations of motion are satisfied, and represent a saddle in the path integral. Therefore the Schwarzian derivative is invariant under SL(2) symmetry $t \rightarrow (at + b)/(ct +d)$ and vanishes for $t(u) = u$.

Since we are on the disk and have a finite boundary length, it is convenient unwrap the circle by reparametrizing
\be
u_{\text{line}} = \tan \frac{\pi}{\beta} t(u_{\text{circle}})
\ee
in terms of the diff($S^1$) reparametrization $f$ satisfying,
\be
t(u_{\text{circle}} + \beta) = t(u_{\text{circle}}) + \beta \; , \quad t'(u_{\text{circle}}) \geq 0 \; .
\ee
We use $u \equiv u_{\text{circle}}$ everywhere else unless stated otherwise. Because the boundary length 
$\beta$ can be interpreted as the inverse temperature, this makes manifest the finite temperature theory.

Quantum gravitational thermal correlators are found by computing the thermal Schwarzian path integral
\be
\langle \mathcal{O} \rangle_{\beta} \equiv \int_{\mathcal{M}} [\mathcal{D} t] \, \mathcal{O}[t] e^{C 
\int_{0}^{\beta} 
du \, \left \{\tan 
\frac{\pi}{\beta} t, \, u \right \}}
\ee
where the integration space $\mathcal{M} = \text{diff}(S^1)/\text{SL}(2,\mathbb{R})$. Using the composition property of the Schwarzian, $\{t(g(u)), u \} = (g')^2 \{t, g \} + \{t, u \}$, this becomes 
\be 
S[t] = -C \int du \, \left \{\tan \frac{\pi}{\beta} t(u), \, u \right \} = \frac{C}{2} \int du \, \left [  
\left(\frac{t''}{t'}\right)^2 
- \left( \frac{2\pi}{\beta} 
\right)^2 (t')^2 \right] \; .
\ee

\subsection{Schwarzian fluctuation theory} \label{app:SchwarzFluc}
To obtain the linearized theory we proceed perturbatively around the saddle by writing $t(u) = u + \epsilon 
k(u)$ and expanding the action in $\epsilon$. To simplify things, let $\beta = 2\pi$; to recover the original $u$ variable rescale $u \rightarrow (2\pi / \beta) u$. After expanding and integrating by parts, the linearized action is
\be \label{eq:EffSchwarzAction}
S_{\epsilon}[t] = \frac{C}{2} \int_{0}^{2\pi} du \, \left[ (k'')^2 -  (k')^2 \right] \; .
\ee
To quantize the fluctuation theory we invert the kernel in the above action by discrete Fourier 
transforming via $\epsilon(u) = \sum_{n} \epsilon_{n} e^{i n u}$ with $n \in \mathbbm{Z}$ and $u \in [0, 2\pi]$. This gives
\be 
S_{\epsilon}[t] = \frac{C}{2} \sum_{n \in \mathbbm{Z}} \, (n^4 - n^2) \epsilon_{n} \epsilon_{-n} \; .
\ee
Inverting, we obtain the fluctuation propagator\footnote{Recall the integration space is modded out by 
gauge redundancy 
SL(2,$\mathbbm{R}$) 

so we can fix the redundant  SL(2,$\mathbbm{R}$) singular modes $n=0, \pm 1$ in the sum and ignore 
when inverting.}
\begin{align*} \label{eq:flucprop}
\langle \epsilon(u_1) \epsilon(u_2) \rangle &= \frac{1}{2\pi C} \sum_{n \neq 0, \pm 1} \frac{e^{i n 
\mathfrak{u}}}{n^2(n^2 - 1)} \\
&= \frac{1}{2\pi C} \left( 1 + \frac{\pi^2}{6} - \frac{(|\mathfrak{u}| - \pi)^2}{2} + (|\mathfrak{u}|-\pi)\sin 
|\mathfrak{u}| 
+ \frac{5}{2} \cos 
|\mathfrak{u}| 
\right) \; , \\
\langle \epsilon(\tau_1) \epsilon'(\tau_2) \rangle &= \frac{\text{sgn}(\mathfrak{u})}{2\pi C} \left( (\pi - 
|\mathfrak{u}| 
) (1-\cos 
|\mathfrak{u}| 
- \frac{3}{2} \sin |\mathfrak{u}| \right) \; , \\
\langle \epsilon'(\tau_1) \epsilon'(\tau_2) \rangle &= \frac{1}{2\pi C} \left( 1 + \frac{1}{2} \cos |\mathfrak{u}| 
- (\pi - |\mathfrak{u}|) \sin 
|\mathfrak{u}| 
 \right) \; , \numberthis
\end{align*}
where $\mathfrak{u} \equiv \tau_1 - \tau_2$.

\subsection{Schwarzian correlation functions} \label{sec:SchCorrs}

To calculate gravitational correlators we need to know how the Schwarzian action (\ref{eq:SchwarzAction}) 
couples to matter correlators. 

As discussed in Section \ref{sec:CFT2AdS2}, we know how to quantify any $n$-point boundary correlator 
$\tilde{\CO}(u_1,\dots,u_n)$  of a bulk CFT$_2$. In 1d reparametrization invariance and conformal 
invariance are the same. Under a conformal transformation $\tilde{\CO}$ transforms as 
\be \label{eq:ConfTrans}
\tilde{\CO}(u_1, \dots , u_n) =  (t'_1)^{\Delta_1/2} \dots (t'_n)^{\Delta_n/2} \tilde{\CO}(t_1(u_1), \dots , 
t_n(u_n)) \; ,
\ee
where $t'_i \equiv d t_i(u_i)/d u_i$. 
Subsequently, from the Schwarzian path integral we can compute gravitational correlators 
\cite{Maldacena:2016upp, 
Blommaert:2019hjr} by computing the quantity
\be \label{eq:schwarzPI}
\< \tilde{\CO} \>_t \equiv \int [\mathcal{D} t] \, \tilde{\CO}(t_i) \, e^{-S[t]} 
\ee
after performing a conformal transformation (\ref{eq:ConfTrans}) on $\tilde{\CO}$.

To perform the path integral we expand around the saddle by letting $t(u) = u + \epsilon k(u)$ with 
$\epsilon$ small. Expanding the action sends $S[t] \rightarrow S_{\text{eff}}[k]$ and expanding
$\tilde{\CO}(t_i)$ to second order in $\epsilon$ lets us use the fluctuation propagator (\ref{eq:flucprop}) to 
obtain the full expression for the gravitational correlator $\< \tilde{\CO} \>_t$.

\subsection{Two-point functions}
Now that know how to calculate gravitational boundary correlators using the Schwarzian, all that remains 
is doing it. Here we review simple examples from the literature.   

Taking a bulk minimal model two-point function of the operator $\CO$ to the boundary gives
\be
G_{2}(t_1, t_2) =
\< \tilde{\CO}_{\tilde{\Delta}}(t_1) \tilde{\CO}_{\tilde{\Delta}}(t_2) \>
 = \frac{1}{t_{12}^{2\tilde{\Delta}}} \; ,
\ee
where $\tilde{\CO}$ is the dual CFT$_1$ operator to $\CO$ consistent with the boundary conditions. 
Under a conformal transformation this becomes
\be \label{eq:2ptconftrans}
\frac{1}{t_{12}^{2\tilde{\Delta}}} \longrightarrow \left(\frac{(f'(t_1) f'(t_2))}{(f(t_1) - f(t_2))^2} 
\right)^{\tilde{\Delta}} \; . 
\ee
To calculate the gravitational contribution to the two-point function, expand (\ref{eq:2ptconftrans}) around 
the saddle $f(t) = \tan \frac{t + \epsilon(t)}{2}$ to second order in $\epsilon$, giving
\be \label{eq:2ptdualexpanded}
G_{2}(t_1, t_2) = \frac{1 + \epsilon \mathcal{B}_1(t_{12}) + \epsilon^2 \mathcal{B}_2(t_{12}) + 
O(\epsilon^3)}{\left( 2 \sin 
\frac{t_{12}}{2} 
\right)^{2\Delta}} \; ,
\ee
where
\begin{align*}
\mathcal{B}_1(t_{12}) &= \tilde{\Delta} \left( k'_1 + k'_2 -  \frac{k_1- k_2}{\tan \frac{t_{12}}{2}} \right) \; , \\
\mathcal{B}_2(t_{12}) &= \frac{\tilde{\Delta}}{4} \left( \frac{k^2_{12}}{(\sin \frac{t_{12}}{2})^2} - 2(k'^2_1 + 
k'^2_2) \right)
+ \frac{\tilde{\Delta}^2}{2} \left( k'_1 + k'_2  - \frac{k_{12}}{\tan \frac{t_{12}}{2} }   \right)^2 \; . \numberthis
\end{align*}
To compute the gravitational correlator substitute (\ref{eq:2ptdualexpanded}) into the effective 
Schwarzian path integral with action (\ref{eq:EffSchwarzAction}), 
\be
\< G_{2}(t_1, t_2) \>_{f} = \frac{1 +\epsilon^2 \< \mathcal{B}_2 (t_{12}) \>}{t^{2\tilde{\Delta}}_{12}} \; ,
\ee
where the linear term in $\epsilon$ vanishes. The final expression involves contracting factors of the  
propagator using (\ref{eq:flucprop}) to give
\begin{align*} \label{eq:twopointgrav}
\< G_{2}(t_1, t_2) \>_{f} &= \frac{1}{(2 \sin \frac{t_{12}}{2})^{2\tilde{\Delta}}} \bigg(1 +
 \tilde{\Delta} \frac{ 2(1-\pi t_{12}) + t^2_{12} - 2 \cos t_{12} + 2 \sin t_{12}(\pi - t_{12} ) }{8 \pi C \sin^2 
\frac{t_{12}}{2} }  \\
&- \tilde{\Delta}^2 \frac{ \big(2 + (2\pi - t_{12})\cot \frac{t_{12}}{2} \big) \big(-2 + t_{12} \cot \frac{t_{12}}{2} 
\big) }{4\pi C} \bigg) \; ,
\numberthis
\end{align*}
reproducing the result of \cite{Maldacena:2016upp}. Factorized four-point functions are a straightforward 
extension of the two-point function and are also treated in the aforementioned reference.

\subsection{Three-point functions}
Three-point functions follow a similar pattern to the two-point ones above. First start with the dual 
CFT$_1$ boundary three-point function
\be
G_{3}(t_1, t_2, t_3) =
\< \tilde{\CO}_{\tilde{\Delta}}(t_1) \tilde{\CO}_{\tilde{\Delta}}(t_2) \tilde{\CO}_{\tilde{\Delta}}(t_3) \>
 = \frac{N_3}{(t_{12} t_{23} t_{13})^{\tilde{\Delta}}} \; ,
 \ee
 where $N_3$ is a normalization constant. Under a conformal transformation this becomes
\be
G_{3}(t_1, t_2, t_3) \longrightarrow \left( \frac{f'_1 f'_2 f'_3}{f_{12} f_{23} f_{13}}  \right)^{\Delta} \; .
\ee
Reparametrizing and expanding around the saddle gives
\be
G_3(t_1, t_2, t_3) = N_3 \frac{1 + \epsilon C_{1}(t_{123}) + \epsilon^2 C_{2}(t_{123}) + O(\epsilon^3) 
}{(t_{12} 
t_{23} 
t_{13})^{\tilde{\Delta}}}
\ee
where
\begin{align*} \label{eq:gravthreepointprops}
C_{1}(t_{123}) &= \tilde{\Delta }\left(
k'_1 + k'_2 + k'_3 
- \frac{k_{12}}{2 \tan \frac{t_{12}}{2}}
- \frac{k_{3}}{2 \tan \frac{t_{23}}{2}}
- \frac{k_{13}}{2 \tan \frac{t_{13}}{2}}
\right) \\
C_{2}(t_{123}) &= 
\frac{\tilde{\Delta}}{8} \left(
\frac{k^2_{12}}{\sin^2 \frac{t_{12}}{2} } 
+ \frac{k^2_{23}}{\sin^2 \frac{t_{23}}{2} } 
+ \frac{k^2_{13}}{\sin^2 \frac{t_{13}}{2} } 
-4(k'_1 + k'_2 + k'_3)
  \right) \\
&+ \frac{\tilde{\Delta}^2}{2} \left(
(k'_1 + k'_2 + k'_3) 
+ \frac{k_{21} \cos t_{12}}{2 \sin \frac{t_{12}}{2} \sin \frac{t_{23}}{2} \sin \frac{t_{13}}{2} } 
+ \frac{k_{13} \cos t_{13}}{2 \sin \frac{t_{12}}{2} \sin \frac{t_{23}}{2} \sin \frac{t_{13}}{2} } 
+ \frac{k_{32} \cos t_{23}}{2 \sin \frac{t_{12}}{2} \sin \frac{t_{23}}{2} \sin \frac{t_{13}}{2} } 
\right)^{2}
\; . \numberthis
\end{align*}
Therefore the gravitational contribution to the three-point function is given by 
\be
\< G_{3}(t_1, t_2, t_3) \>_{f} = N_3 \frac{1 + \epsilon^2 \<C_{2}(t_{123})\> + O(\epsilon^3) }{(t_{12} t_{23} 
t_{13})^{\tilde{\Delta}}} \; ,
\ee
where again the $\epsilon$-propagator terms are contracted using (\ref{eq:flucprop}), reproducing \cite{Qi:2018rqm}.

\bibliographystyle{ieeetr}
\bibliography{CFT2_Bulk_Gravity}{}

\end{document}